\documentclass[10pt,twocolumn]{article}

\usepackage[a4paper, margin=0.75in, columnsep=0.3in]{geometry}
\usepackage[utf8]{inputenc}
\usepackage[T1]{fontenc}
\usepackage{lmodern}

\usepackage{cite}
\usepackage{amsmath, amssymb, amsfonts}
\usepackage{graphicx}
\graphicspath{{figures/}}
\usepackage{textcomp}
\usepackage{url}
\usepackage{array}
\usepackage{bm}
\usepackage{booktabs}
\usepackage{multirow}
\usepackage{xcolor}

\usepackage{microtype}
\usepackage[font=footnotesize,
            labelfont=bf,
            labelsep=period,
            justification=justified,
            singlelinecheck=false]{caption}

\usepackage{etoolbox}
\AtBeginEnvironment{table}{\footnotesize\setlength{\tabcolsep}{3pt}}
\AtBeginEnvironment{table*}{\footnotesize\setlength{\tabcolsep}{3pt}}

\definecolor{olivelink}{RGB}{85,107,47}      
\usepackage[colorlinks=true,
            citecolor=olivelink,
            linkcolor=olivelink,
            urlcolor=olivelink,
            filecolor=olivelink]{hyperref}

\newcommand{\history}[1]{}
\newcommand{\doi}[1]{}
\newcommand{\corresp}[1]{}
\newcommand{\tfootnote}[1]{}
\renewcommand{\markboth}[2]{}

\newlength{\titlepgskip}
\newcommand{\address}[2][]{}

\newenvironment{keywords}
  {\medskip\noindent\textbf{Keywords --} }{\par\medskip}

\renewcommand{\thesection}{\Roman{section}}
\renewcommand{\thesubsection}{\Roman{section}-\Alph{subsection}}

\title{Contextual Geospatial Features for Identifying Informal
Environmental-Health Hazards Undetectable from Satellites:
A ULAB Case Study}

\author{
  Naia Ormaza-Zulueta$^{1,2,*}$
  \and
  Zia Mehrabi$^{2,3}$
  \\[0.6em]
  \small $^{1}$Institute for Resources, Environment and Sustainability,
         University of British Columbia, Vancouver, BC V6T 1Z4, Canada \\
  \small $^{2}$Better Planet Laboratory, University of Colorado
         Boulder, Boulder, CO 80309, USA \\
  \small $^{3}$Department of Environmental Studies, University of
         Colorado Boulder, Boulder, CO 80309, USA \\[0.4em]
    \small \textsuperscript{*}Corresponding author: naia.ormazazulueta@ubc.ca
}

\date{}

\begin{document}

\twocolumn[\begingroup
  \maketitle
  \begin{@twocolumnfalse}

\begin{abstract}
Reliable, scalable detection of informal, small-scale environmental-health hazards (used lead-acid battery (ULAB) recycling, household-scale e-waste burning, indoor mercury amalgamation, brick kilns, small tanneries) remains an unsolved problem. These operations are invisible to satellites and absent from formal registries, yet disproportionately harm low-income populations in low- and middle-income countries (LMICs). This paper articulates the problem class and explores a possible response: contextual geospatial features (point-of-interest text, infrastructure proximity, socio-economic structure), with case-specific feature design informed by domain expertise. We use ULAB recycling as a demonstration case, drawing on 164 verified sites in Bangladesh and India from Pure Earth's Toxic Sites Identification Programme (TSIP). At this sample size, five-fold cross-validation on the training set cannot statistically distinguish the engineered contextual features from a simple two-feature socio-demographic baseline. The added value only becomes visible when we evaluate outside the training set. On 172 held-out informal-recycling sites in non-NCR India and Bangladesh, the model assigns scores several times higher than to matched random urban controls; on a zero-shot scan of India's National Capital Region, it flags a substantially more selective set of high-risk cells than the open-data baseline; and on an independent set of 131 regulatory-confirmed formal recyclers, informal sites score materially higher than formal ones in non-NCR India, indicating that the model is picking up informal-recycler-specific structure rather than generic industrial signal. We frame these results as exploratory rather than confirmatory: label sparsity, gaps in point-of-interest coverage, and untested transfer beyond South Asia all remain open. We close with seven open problems and invite the environmental-health and geospatial machine-learning communities to engage with informal-hazard detection as a class of problems worth solving.
\end{abstract}


  \vspace{1.5em}
  \end{@twocolumnfalse}
\endgroup]

\section{Introduction}
\label{sec:intro}

Environmental remote sensing has matured to the point where many important global hazards can be located, classified, and monitored from space. Global-scale datasets of mining operations, derived from expert visual interpretation of high-resolution satellite imagery, catalogue mining and processing sites at country and regional scale \cite{ref1}. Sentinel-5P TROPOMI tracks daily SO$_2$, NO$_2$, and methane plumes from large industrial sources \cite{ref2}. Hansen et al.'s multi-temporal Landsat product detects forest loss at 30 m globally \cite{ref3}. VIIRS night-time radiance enumerates oil-and-gas flares to the individual stack \cite{ref4}. SAR backscatter detects oil-spill slicks, and Amazon Mining Watch have shown that even some informal hazards with discrete surface footprints (open-pit artisanal small-scale gold mining, ASGM) can be mapped at scale from optical imagery \cite{ref30}. The common architecture of these successes is the same: the hazard has a direct, spatially-resolvable physical signature, larger than the sensor's footprint and spectrally distinct from its surroundings, that can be observed from space.

A second class of environmental-health hazards does not satisfy this premise. Informal small-scale operations, such as used lead-acid battery (ULAB) recycling, informal e-waste burning, household-scale or indoor ASGM mercury amalgamation, backyard brick kilns, illegal dumping in dense urban areas, small tanneries and more, share four properties that defeat direct satellite detection : (i) operational signatures below the resolution of even the highest resolution imagery (signatures within a single household courtyard to small workshop, and pollutants emitted in small doses); (ii) embeddedness in residential or mixed-use landscapes that dominate the spectral neighbourhood; (iii) absence of distinctive large-scale physical signatures (no slag pile, no chimney, no clearing); and (iv) absence from formal registries or permits, which removes the cheap label-acquisition shortcut available for formal facilities. The result is a structural gap in the current environmental remote sensing toolkit: an entire category of public health hazards that disproportionately harm low-income populations and that are largely invisible to the methods based only on current satellite data alone. This class of features, and the task of identifying informal environmental hazards, represents a large unsolved problem.

This paper highlights this problem and explores potential ways that it may be overcome, using domain transfer from other fields, such as ecology. We show the potential of inferring the hazard's location from the contextual geospatial structure of the places where it operates, rather than from its physical signature. This is built up from the idea that informal environmental-health hazards do not appear at random across the landscape -- they cluster where their material inputs accumulate, where waste-management infrastructure is weak, and where socio-economic conditions favour informal labour. 

We assess that translating this intuition into a readily available detection pipeline requires three jointly necessary ingredients: (a) public geospatial data dense enough to expose the relevant local patterns; (b) case-specific epidemiological knowledge of which POI (point-of-interest) types, business names, and infrastructure features co-occur with the target hazard, and how they should be combined; and (c) some quantity of verified positive and negative locations for fitting and testing. We posit these ingredients can in principle be translated to any currently undetectable hazards, but the feature design is itself a domain-knowledge task that must be redone for each hazard type.

We instantiate these ingredients for ULAB recycling, as an example case. ULAB recycling is chosen because of its public-health severity and because a moderate volume of field-verified, georeferenced sites exists in Pure Earth's Toxic Sites Identification Programme (TSIP), to support initial quantitative model fitting and external validation. A joint UNICEF (United Nations Children's Fund)--Pure Earth report estimates that one in three children globally carries blood lead concentrations above 5 $\mu$g/dL, the threshold associated with irreversible neurological harm \cite{ref11}. In LMICs (low- and middle-income countries), a substantial and under-documented fraction of this burden originates from informal ULAB recycling, where batteries are broken open, drained, and smelted without pollution controls, often within residential neighbourhoods \cite{ref12}. These sites are overwhelmingly informal: they do not appear in business registries, environmental permits, or industrial databases, and they are similarly opaque in trade data, so even national totals (let alone subnational spatial distributions) are estimated rather than observed. Pure Earth's TSIP, the largest georeferenced inventory, has catalogued several thousand sites across more than 50 countries \cite{ref7}, but this represents a small fraction of the true total. The traditional approach to locating contamination sources relies on field teams with portable X-ray fluorescence (XRF) analysers: slow, expensive, and impossible to scale to national coverage \cite{ref8}. Blood-lead surveillance detects exposed populations but cannot resolve point sources \cite{ref13}.

We support our analysis by the knowledge that lead acid batteries accumulate at auto-repair shops, scrap dealers, motorcycle workshops and battery resellers, all of which can be mapped with varying completeness in OpenStreetMap, Overture Maps, and Google Places. The relevant socio-economic context (low wealth indices, dense peri-urban occupation) in turn can be captured by geospatial surfaces, such as Meta's Relative Wealth Index \cite{ref5} and WorldPop population density \cite{ref6}. And ULAB-specific knowledge (which keywords to match, which POI categories to weight, which transliterations to include) can easily developed iteratively from the literature, field-investigation reports, and with validation by Pure Earth and Toxics Link experts and leading experts in the field \cite{ref7, ref8, ref9}. 

The contribution of this paper is to present the unsolved problem, and use ULAB as a demonstration case for how this problem class may be approached. It is a characterisation of the problem class; an exploratory demonstration that contextual features carry signal across multiple testing tiers external to the training set; and a roadmap of open problems applicable to the broader class of below-resolution informal hazards.

\subsection{Related work}
\label{sec:relwork}

The idea of using contextual features to build probability surfaces for objects not otherwise detectable is widely used in a number of fields on which we build our idea. Perhaps the most prominent case is that of ecology, where species can rarely be `seen' or detected at scale, and so populations are instead predicted from a range of climatic, topological, edaphic, and other features for which they co-occur \cite{ref-elith}. Other examples include estimates of poverty driven by key features associated with that poverty, like the frequency of mobile handset ownership in a given neighborhood \cite{ref5}, use of social media geolocated and time-stamped interaction frequencies, or point-of-interest data, alongside satellite data to classify urban areas, into whether they are industrial, residential, etc. \cite{ref16, ref-liu-urban}. Closer to our setting, machine-learning approaches have been used to detect informal settlements in developing countries from low-resolution multispectral imagery \cite{ref17}, demonstrating that contextual and below-resolution detection is tractable in adjacent problem classes. Similarly epidemiologists have recognized the value of using contextual data and digital `traces' of human activity to infer disease spatial and temporal dynamics \cite{ref-salathe-digital-epi}. A well established research program, the Institute for Health Metrics and Evaluation (IHME) \cite{ref13, ref-ihme-gbd}, while not taking as granular an approach to define context, does use a range of geospatial covariates to downscale and develop small area estimates for a range of health indicators, from prevalence of breastfeeding and child malnutrition to incidence of tropical parasites. Our ideas draw on these ideas but build on them specifically to address the problem of environmental hazards. The present paper is positioned, to the authors' knowledge, as among the first explicit attempts to fuse POI-text matching with case-specific epidemiology, socio-demographic priors, and infrastructure features for informal-hazard detection, and to characterise (rather than claim to solve) the failure modes that emerge.

The five specific contributions of this paper are:

(i) Framing the detection of informal environmental-health hazards as a methodological problem class characterised by an operational scale below current sensor resolution, residential embeddedness, absence from formal registries, and structural invisibility to methods built for remote-sensed formal smelters, plumes, deforestation, or gas flares, due to the absence of any distinctive physical signature that imagery alone can pick up.

(ii) Proposal of contextual geospatial features (POI text, infrastructure proximity, socioeconomic structure) as a methodological alternative for environmental-health hazards, and stress-testing the proposal on used lead-acid battery (ULAB) recycling, a high-stakes instance for which a low inventory of field-verified sites exists.

(iii) Training demonstration models and evaluating the potential benefits of different configurations of features in aggregate for predicting ULAB sites in Bangladesh and India, for example using proprietary vs public contextual POI data compared to baseline models of population and wealth, for high resolution geolocating of hazard sites Alongside external validation excercises.

(iv) Demonstrating why specific contextual features matter for hazard site identification, including geospatial feature interactions, through model explainability methods such as SHAP (SHapley Additive exPlanations) attribution, visualization, and distributional shifts.

(v) Articulating seven open research questions for informal environmental-health hazards, based on learnings from ULAB and applicable to the broader class of below-resolution informal hazards (informal e-waste burning, artisanal small-scale mining, backyard brick kilns, illegal dumping); and enumerating key threats to validity for researchers to 'watch' for in addressing this unsolved problem.

\section{Why ULAB Detection is a Hard Problem}
\label{sec:hard}

Informal ULAB recycling presents a detection challenge fundamentally different from canonical satellite remote-sensing tasks. Four structural characteristics make it particularly difficult; each generalises to the broader informal environmental hazard class introduced in Section~\ref{sec:intro}.

\subsection{Invisibility in imagery and registries}
\label{sec:hard-invis}
Informal ULAB operations occupy areas from a single-home courtyard to a small workshop, spectrally indistinguishable from surrounding residential or light-commercial land use. They also appear in no official database (like toxic release inventories which formal sites may), and are similarly opaque in trade and customs statistics: ULAB flows pass through informal scrap supply chains that go unrecorded in national imports, exports, and recycling tonnages, so even country-level estimates are inferred rather than observed.

\subsection{Extreme scarcity of labeled data}
\label{sec:hard-scarcity}
Pure Earth's TSIP records yielded 50 verified ULAB positives\footnote{TSIP's database holds a substantially larger pool of georeferenced ULAB records than the 50 retained here. The 50 are those whose coordinates were manually cross-checked by the lead author. The verification protocol is described in Section~\ref{sec:methods-data}, and the larger unverified TSIP pool is reserved for area-based external validation (Section~\ref{sec:results-nbhd}).} in Bangladesh and India at modern GPS (Global Positioning System) precision. Label scarcity is itself a defining feature of the problem class and generating new labels through human-in-the-loop learning is part of the open challenges this paper articulates. Building a credible negative class is similarly demanding: confirming that a location does not host ULAB activity requires either field verification or confidence in survey completeness, neither of which is broadly available outside dedicated field campaigns. 

\subsection{Geographic heterogeneity}
\label{sec:hard-geo}
\begin{figure}[!t]
\centering
\includegraphics[width=\columnwidth,keepaspectratio]{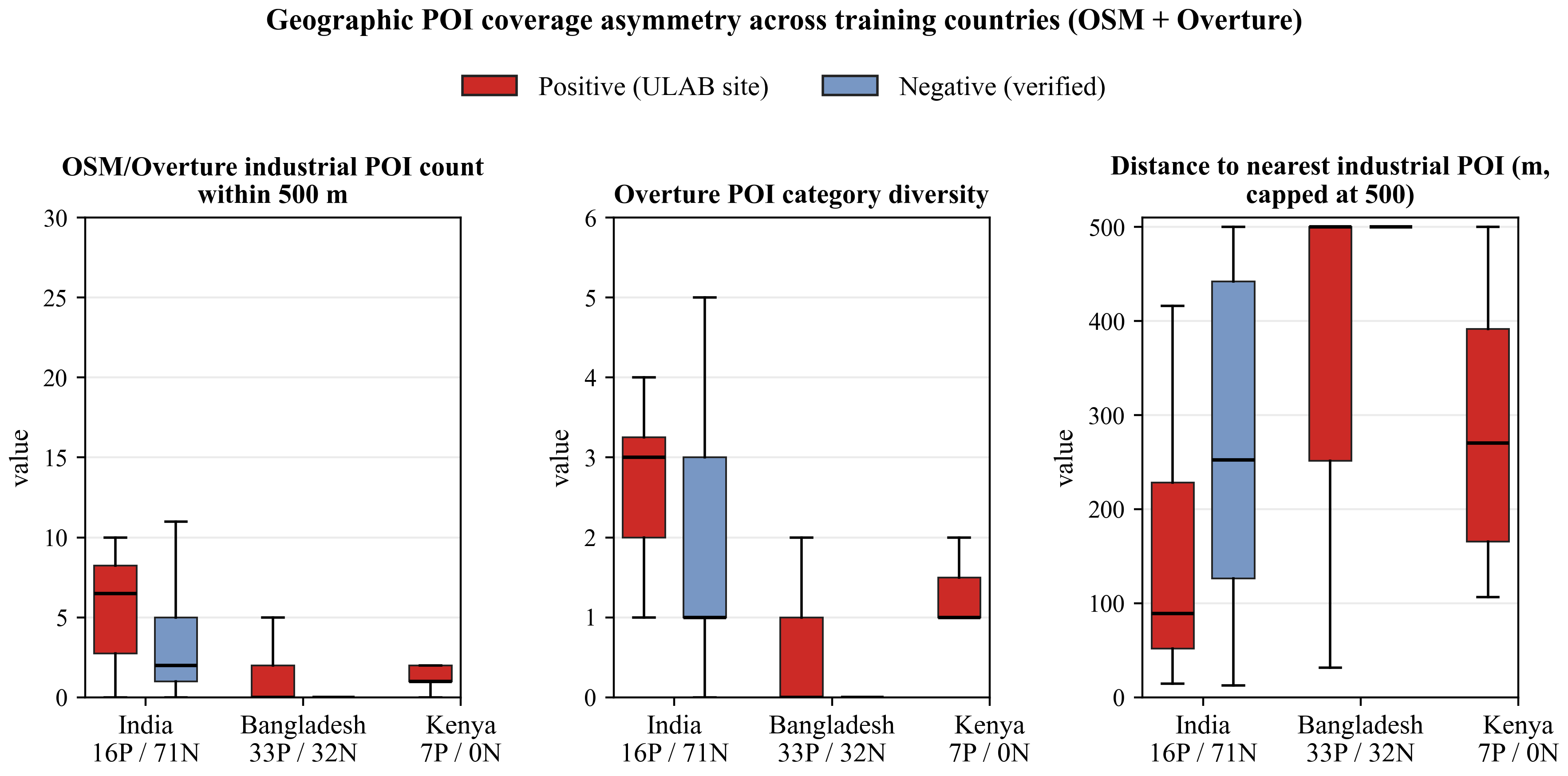}
\caption{ULAB geographic POI coverage asymmetry across countries used in demonstration model, using the Open POI features (OSM and Overture). India has the densest POI ecology (highest counts within 500 m, deepest category diversity). Bangladesh has sparse Overture coverage. This asymmetry is a precondition for the cross-country failure modes documented in Section~\ref{sec:results-transfer}.}
\label{fig1}
\end{figure}

The morphology of informal recycling varies across contexts. In Bangladesh, operations concentrate in dense peri-urban clusters along transport corridors, often adjacent to automotive markets. In India, they are more diffusely distributed across urban and semi-rural areas. A model trained predominantly on South Asian morphology may learn feature associations (density of auto-repair shops, proximity to highways, population thresholds) that do not transfer to other cities, or regions such as sub-Saharan Africa or Latin America. POI data is also geographically uneven: OpenStreetMap and Google Places have markedly sparser coverage in rural sub-Saharan Africa than in South Asia, for example \cite{ref14, ref15} (Fig.~\ref{fig1}).

\subsection{Class imbalance at deployment scale}
\label{sec:hard-imbalance}
Even in countries with high ULAB prevalence, contaminated sites occupy a vanishing fraction of the landscape. Within bounded urban scans the base rate of TSIP-documented sites against the population of candidate cells is on the order of $10^{-3}$ at the half-kilometre cell sizes used for deployment, and falls towards $10^{-4}$ at the finer resolutions that facility-level localisation would require. Pure Earth's Toxic Site Identification Programme (TSIP), the key dataset for ULAB site locations, catalogues only sites that have entered its database through field campaigns and is not systematic, so the true informal-recycling distribution is plausibly larger but currently unobserved. Even so, at the deployment scale any classifier must operate with high precision to produce actionable candidate lists for field teams to investigate, rather than thousands of false positives \cite{ref21}. The specific NCR deployment scan that anchors the per-cell numbers in this paper is detailed in Section~\ref{sec:results-ncr}. 

\section{Methods Given Limited Data}
\label{sec:methods}

\begin{figure}[!t]
\centering
\includegraphics[width=\columnwidth,keepaspectratio]{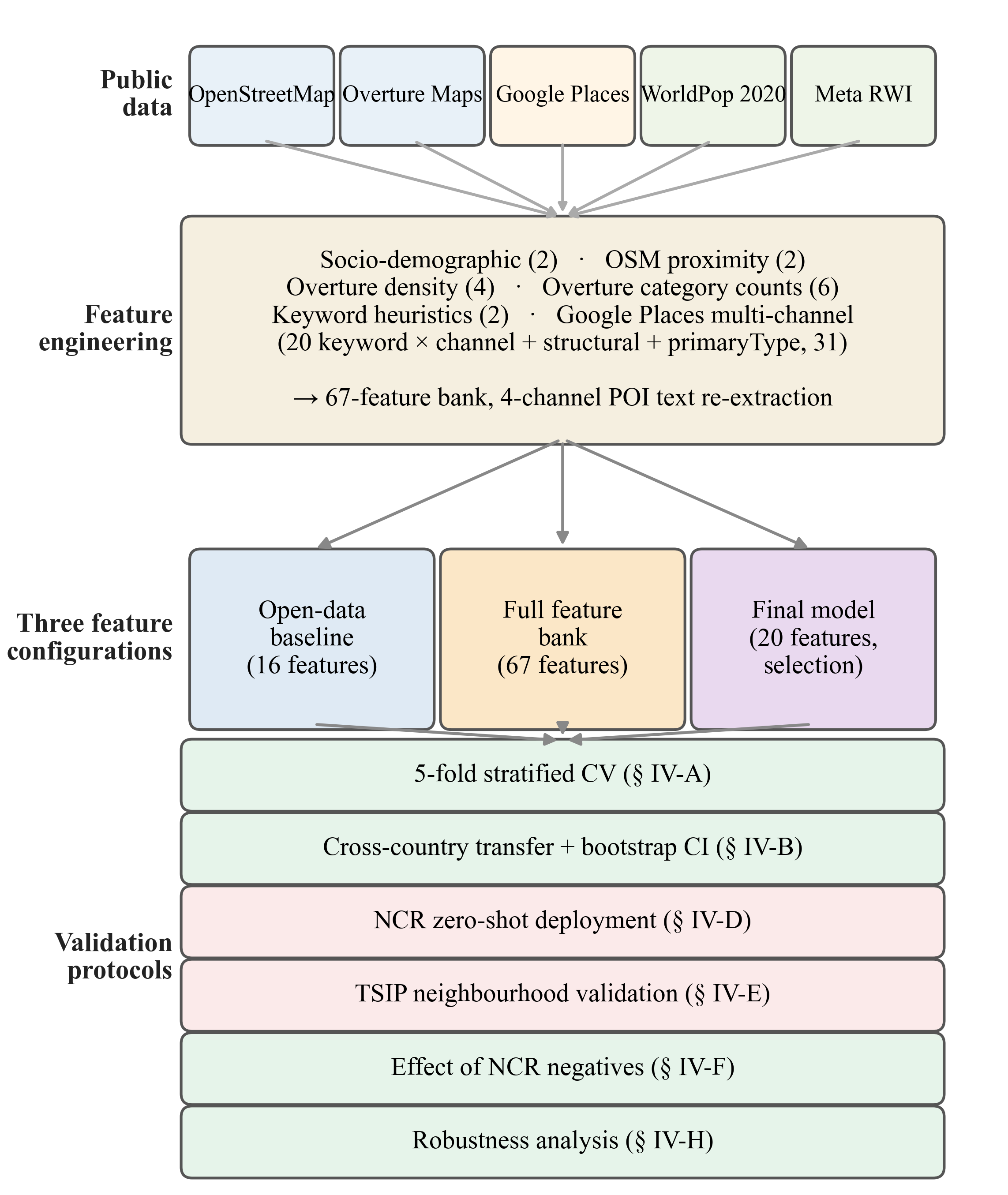}
\caption{Pipeline overview for demonstration model. Five public data sources (OpenStreetMap, Overture Maps, Google Places, WorldPop, Meta Relative Wealth Index) feed a 67-candidate feature bank; three XGBoost configurations (open-data baseline, full feature bank, 20-features selected by univariate discrimination and selection-stability across the 5-fold splits) are trained on the same sparse n = 164 label set; five validation protocols (5-fold CV + LOOCV, BGD$\leftrightarrow$IND cross-country LOCO (leave-one-country-out), India NCR zero-shot deployment, area-based neighbourhood validation against TSIP ground truth, full 20-feature out-of-region transfer) probe training-set discrimination, transfer to held-out testing tiers, and deployment behaviour.}
\label{fig2}
\end{figure}

Given the constraints above, this paper implements a simple tabular gradient-boosted classifier with candidate and target relevant geospatial features. It is not the intent to solve the ULAB prediction in this paper, especially given the very small sample size, but to use a demonstration model and evaluation of it, to assist in understanding how the problem of informal environmental hazards may be approached. The full pipeline is summarised in Fig.~\ref{fig2}.

\subsection{Training data}
\label{sec:methods-data}

Positives (n = 50) were drawn from Pure Earth's TSIP database for Bangladesh and India, each location was manually verified against high-resolution Google Maps, Street View, Overture, and satellite-imagery overlays, and found to land at or within metres of visible facility infrastructure consistent with the TSIP site description. Records whose coordinates fell on empty fields, water, or otherwise inconsistent land cover, and were not recent enough, were excluded from training. A larger TSIP pool (n=172) carrying Pure Earth's documented ULAB coordinates was set aside for external validation at coarser scales (see Section~\ref{sec:results-nbhd} ) rather than training (which would propagate any unvalidated per-record coordinate spatial error). 

Negatives (n = 72) were sampled from urban areas in India and Bangladesh with no documented ULAB activity, each cross-checked by the lead author against high-resolution Google Maps, Street View, Overture, and satellite imagery to rule out visible ULAB-consistent infrastructure. The negative pool was then augmented with three further tiers: 11 verified deployment-region negatives within India's NCR (Section~\ref{sec:results-ncr}); 6 TSIP-confirmed-formal lead smelters in India (used as informative ``hard'' negatives that look industrial but are licensed facilities, not informal recyclers); and 25 domain-rule synthetic negatives -- cells drawn programmatically from three rule-defined strata of Bangladesh and India that are very unlikely to host informal ULAB activity: (i) wealthy-urban cells (top RWI decile with low industrial-POI density), (ii) rural low-population cells (population density below the urban threshold and no roads classified as ``residential'' or higher in OSM), and (iii) remote cells (more than 30 km from the nearest settlement of $>$ 5{,}000 inhabitants). These rule-based negatives broaden the negative class beyond the manually-reviewed urban pool without requiring additional field verification. The final training set used for all results for this demonstration in Section~\ref{sec:results} is therefore n = 164: 50 positives (33 Bangladesh + 17 India) and 114 negatives (72 manually-reviewed urban + 11 NCR + 6 TSIP-formal + 25 synthetic).

\subsection{Feature engineering}
\label{sec:methods-features}

For each location a candidate feature bank is computed from five publicly available sources, grouped into the six semantic channels listed in Table~\ref{tabI}: POI features are grouped into two sets of analytic channels by whether they are Open or Proprietary. The distinction is methodologically substantive: the Open POI channels are reproducible at zero marginal cost, while the Proprietary POI channels introduce per-query cost, a single-vendor dependency, and coverage that is geographically uneven in ways that themselves track Google's deployment priorities. Identifying which channel carries the deployment-relevant signal is, therefore, both a model-attribution question and a policy-relevant one about what kinds of data the methodology needs to be reproducible at scale. 

All POI features are complemented with socio-demographic features (\texttt{pop\_density} from WorldPop and \texttt{wealth\_rwi} from Meta's Relative Wealth Index), which also offer baseline. One caveat on the use is that WorldPop and Meta RWI are themselves machine-learning based products that embed urban/commercial structure (e.g. cellphone handsets, road density, building footprints) \cite{ref5, ref6}. Thus the question is largely what POI data add to that signal already embedded in these products.

Briefly the Open POI channels comprise features drawn from openly-licensed features: Open Street Map (OSM) road/water proximity features (OpenStreetMap ODbL) and Overture features capturing industrial-POI density, distance to the nearest industrial POI, POI-category diversity, and keyword-tag flags (Overture Maps Foundation, ODbL/CC-BY \cite{ref27}). 

The Proprietary POI channels are comprised of Google Places features (Google Maps Platform, commercial API key required), built from four text channels per cached place (`displayName', `formattedAddress', `primaryType', and the full `types' list); a single-channel variant that matches only against `displayName' serves as a comparison point in Section~\ref{sec:results-ablation}. The four-channel formulation recovers substantially more keyword matches per site (e.g. ``industrial area'' and ``industrial estate'' tokens in `formattedAddress', and Bengali- and Hindi-transliterated terms in the same field) that the display-name-only matcher misses. The proprietary-POI channel reported throughout this paper uses the four-channel formulation throughout. The Google Places lexicon comprises five tiers (battery-direct, smelting/recycling, metal-work, auto-repair, electrical) of English, Hindi, and (where transliterated) Bengali keywords; kabaad, raddi, bhangaar, lohar, dhalai, batri; plus brand names (Exide, Amaron, Luminous). Bengali coverage is currently narrower than Hindi and is a known limitation; expanding the Bengali lexicon is one of the open problems in Section~\ref{sec:roadmap-poi-text}. Each tier is matched against four text channels (display name, formatted address, primary type, full type list) producing tier $\times$ channel $\times$ any-channel counts. A negative-keyword list (laptop, phone, restaurant, school, ...) suppresses common false matches. Structural counts capture place density, address-token diversity, and primary Type-density panels.

Three feature configurations are then evaluated to disentangle which channels carry signal:

(i) The open-data baseline retains only OSM/Overture POI + WorldPop + Meta RWI features (16 features total). This is the configuration achievable from fully open-licensed data without any proprietary API access.

(ii) The full feature bank (n=67) adds the proprietary POIs, including multi-channel Google Places features and a panel of primaryType densities (auto\_parts\_store, car\_repair, tire\_shop, manufacturer, etc.) and address tokens ("industrial area", "industrial estate").

(iii) A selected set of 20 features, from both Open and Proprietary POIs selected using a univariate discrimination filter (positive/negative mean ratio $\geq$ 1.5) combined with a selection-stability check across the 5-fold splits, keeping 2 socio-demographic, 10 Open POI (e.g. from OSM, Overture density, Overture keyword), and 10 selected Google Places features. We note that ideally this selection process should be done on an independent dataset than that used in downstream models-- although this choice represents the sparsity problem more widely.

\begin{table*}[!t]
\caption{TABLE I. FEATURE GROUPS AND CONFIGURATIONS}
\label{tabI}
\centering
\setlength{\tabcolsep}{4pt}
\footnotesize
\begin{tabular}{p{2.4cm}p{3.0cm}p{6.5cm}ccc}
\hline
\textbf{Group} & \textbf{Source} & \textbf{Features} & \textbf{Open-data} & \textbf{Full bank} & \textbf{Final model} \\
\hline
Socio-demographic & WorldPop 2020; Meta RWI & pop\_density, wealth\_rwi & $\checkmark$ & $\checkmark$ & $\checkmark$ \\
Infrastructure proximity & OpenStreetMap via osmnx \cite{ref22} & dist\_road\_m, dist\_water\_m & $\checkmark$ & $\checkmark$ & $\checkmark$ \\
Overture industrial density & Overture Maps POI & dist\_industrial\_m, industrial\_count\_500m\_log, industrial\_count\_1km\_log, poi\_category\_diversity & $\checkmark$ & $\checkmark$ & $\checkmark$ \\
Overture category counts & Overture Maps POI & battery\_count\_500m, metal\_count\_500m, auto\_count\_500m, scrap\_count\_500m, electric\_count\_500m, industrial\_count\_500m (6 features) & $\checkmark$ & $\checkmark$ & --- \\
Keyword heuristics & Overture primary\_tag & has\_industrial\_tag, is\_high\_risk\_tag & $\checkmark$ & $\checkmark$ & $\checkmark$ \\
Google Places --- multi-channel & Google Places (Nearby Search, 4 text channels) & Five-tier keyword counts $\times$ {displayName, formattedAddress, types, any-channel} = 20 features; structural counts (total places, industrial-area address hits, strong primaryTypes, ULAB score, etc.) and primaryType densities (manufacturer, car\_repair, auto\_parts\_store, tire\_shop, \dots{}) --- 38 features in the full bank, 10 retained after selection & --- & $\checkmark$ & $\checkmark$ (10) \\
Feature count & --- & --- & 16 & 48 & 20 \\
\hline
\end{tabular}
\end{table*}

\subsection{Classifier}
\label{sec:methods-classifier}

A simple XGBoost classifier \cite{ref24} is trained with n\_estimators = 200, max\_depth = 3 (shallow trees), learning\_rate = 0.05, min\_child\_weight = 3, subsample = 0.8, colsample\_bytree = 0.8, L1 regularisation $\alpha$ = 0.1, L2 regularisation $\lambda$ = 1.0. Class imbalance is handled via scale\_pos\_weight = n\_neg / n\_pos $\approx$ 1.46. Generalisation is evaluated via 5-fold stratified cross-validation (StratifiedKFold, shuffle = True, random\_state = 42). For the deployment model, probabilities are calibrated by isotonic regression in nested 5-fold CV \cite{ref25}. As a sanity-check baseline, an L2 logistic regression is also reported (L2-regularised, standardised features, same class-balancing) on the same feature set in Table~\ref{tabII}. The operational deployment threshold is $\tau$ = 0.5 throughout this paper unless otherwise specified.

\subsection{Evaluation strategy}
\label{sec:methods-eval}

It is not the primary purpose of this paper to present a model that we know works at the expected deployment base rate below $10^{-4}$ at the finer resolutions facility-level localisation would require. We do however use a demonstration model to better understand how the contextual modeling approach performs, and what its advantages may be by performing a range of evaluation techniques:

\textit{Training-set rank discrimination} (Section~\ref{sec:results-cv}). Five-fold stratified CV and LOOCV on the n = 164 training set, with model-class robustness checks (XGBoost vs L2 logistic regression) and a 2-feature socio-demographic baseline that exposes how much of the apparent signal comes from contextual priors that any reasonable classifier would exploit. We note this is not our primary evaluation method, given the known sparsity of labels, but we include it for completeness.

\textit{Feature attribution and channel reliance} (Sections~\ref{sec:results-fi} and~\ref{sec:results-ablation}). TreeSHAP \cite{ref28} per-positive attribution on the training set and on held-out testing populations, contrasted with XGBoost-gain rankings and with a nested ablation across the three feature configurations, and a feature-group ablation (Table~\ref{tabIV}) that drops one channel family at a time. The right question at small $n$ is not headline AUC but which channels carry the deployment-relevant signal, at the locations the model needs to find.

\textit{Cross-country generalisation} (Section~\ref{sec:results-transfer}). Standard leave-one-country-out (BGD$\leftrightarrow$IND) and a more deployment-relevant variant: retrain on Bangladesh-only data and predict on held-out India and Kenya positives. This separates training-set sharpness from genuine geographic transfer.

\textit{Deployment calibration and external validation} (Sections~\ref{sec:results-ncr} and~\ref{sec:results-nbhd}). Zero-shot deployment of each trained model to a populated-cell grid in India's NCR, scored against fraction-above-threshold and spatial coherence (not AUC); area-based enrichment of high-score cells in buffers around TSIP-documented informal sites; and a fully independent validation against 131 regulatory-registered formal recyclers, none of which entered any training pool.

A final tier of additional robustness diagnostics (Appendix~\ref{sec:appx-coverage}) probes the methodology for spatial-autocorrelation overestimation, hard-negative trade-offs, and Google Places coverage bias.

\section{What We Found}
\label{sec:results}

We find POI engineered configurations help discriminate ULAB sites, though the gap over a 2-feature socio-demographic baseline is within fold-to-fold variability (Section~\ref{sec:results-cv}). Engineered POI text features tighten the deployment-region score distribution and yield a more selective, spatially coherent risk surface (Section~\ref{sec:results-ncr}). Area-based enrichment around documented informal sites is detectable within the deployment region at the 5\,km scale on peak score (Section~\ref{sec:results-nbhd}). We find evidence of out-of-region transfer to TSIP-informal sites in non-NCR India, while Bangladesh transfer is weaker and depends on the engineered POI channels ( Section~\ref{sec:results-nbhd}). A fully out-of-source validation against 131 regulatory-confirmed formal recyclers (none of which entered any training pool) discriminates formal from informal industrial activity in non-NCR India and in the deployment region (Section~\ref{sec:results-nbhd}, Table~\ref{tabFormal}), though the combined formal-vs-informal test is blurred by the weaker Bangladesh signal. All these findings we read as evidence that a signal exists across these tiers, warranting further investigation of contextual features; results are reported with explicit caveats at each step (Section~\ref{sec:results-threats}). 

\subsection{Five-fold cross-validation on training data}
\label{sec:results-cv}

Our intent in this demonstration is not to train a model fit for production, but to build a model with different specifications that can in turn be used to inspect the \textit{potential} value of contextual features for modeling informal ULAB facilities and for detecting informal environmental hazards more broadly. 

\begin{figure}[!t]
\centering
\includegraphics[width=\columnwidth,keepaspectratio]{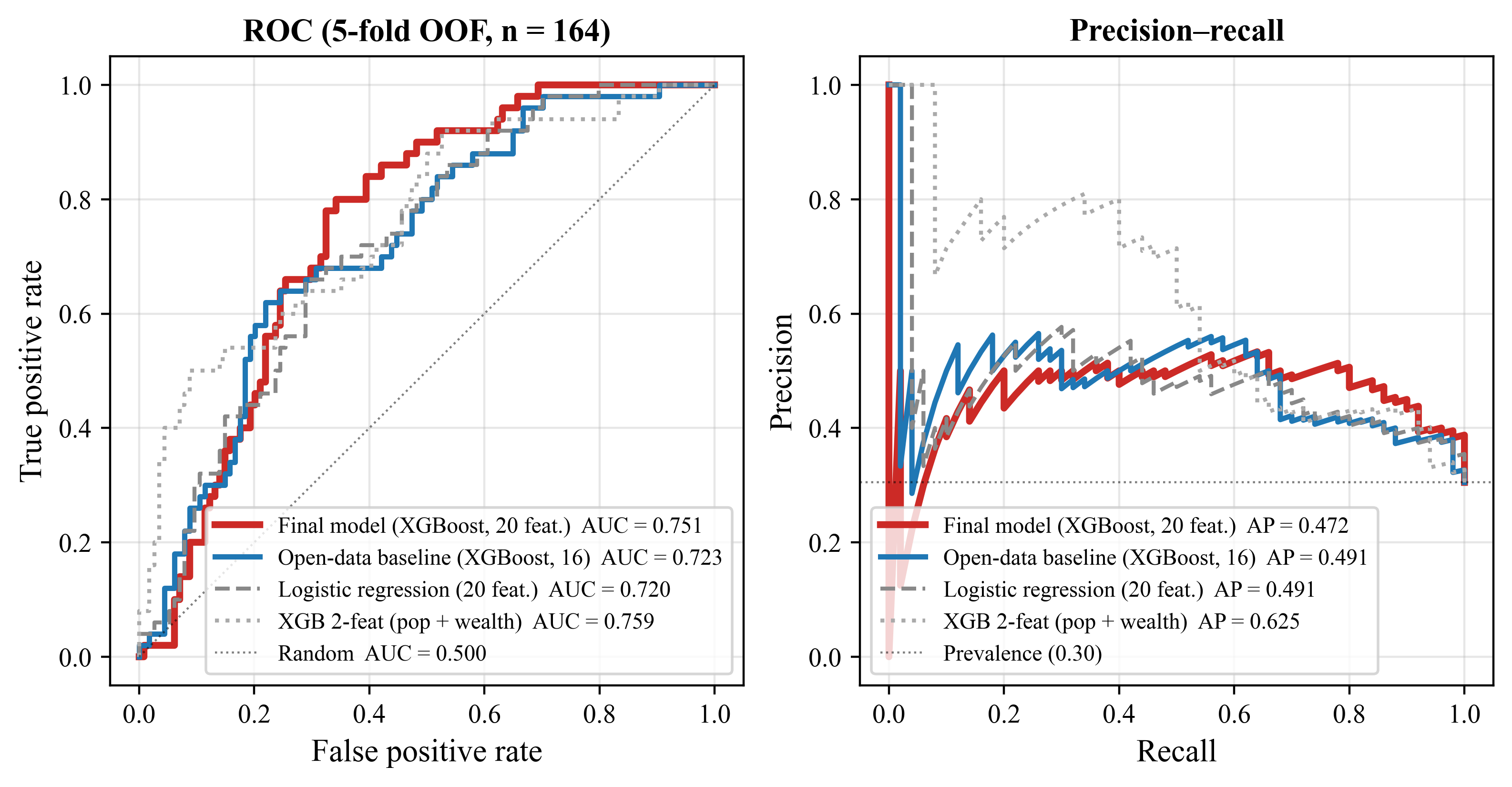}
\caption{5-fold CV ROC (left) and precision-recall (right) curves on the n = 164 unified training set, with weak-baseline reference curves. The XGBoost "Final model" (20 features) and "open-data baseline" (16 features) achieve pooled 5-fold CV AUC $\approx$ 0.77 and 0.72 respectively; an L2 logistic-regression on the same 20-feature panel gives AUC $\approx$ 0.73, and a 2-feature (pop\_density + wealth\_rwi) XGBoost reaches AUC $\approx$ 0.76. The gap between the engineered XGBoost models and the simple baselines is within fold-noise on this training-set metric; the value of Proprietary-POI multi-channel text features becomes apparent on out-of-region transfer (Section~\ref{sec:results-nbhd}) and deployment calibration (Section~\ref{sec:results-ncr}).}
\label{fig3}
\end{figure}

\begin{figure}[!t]
\centering
\includegraphics[width=\columnwidth,keepaspectratio]{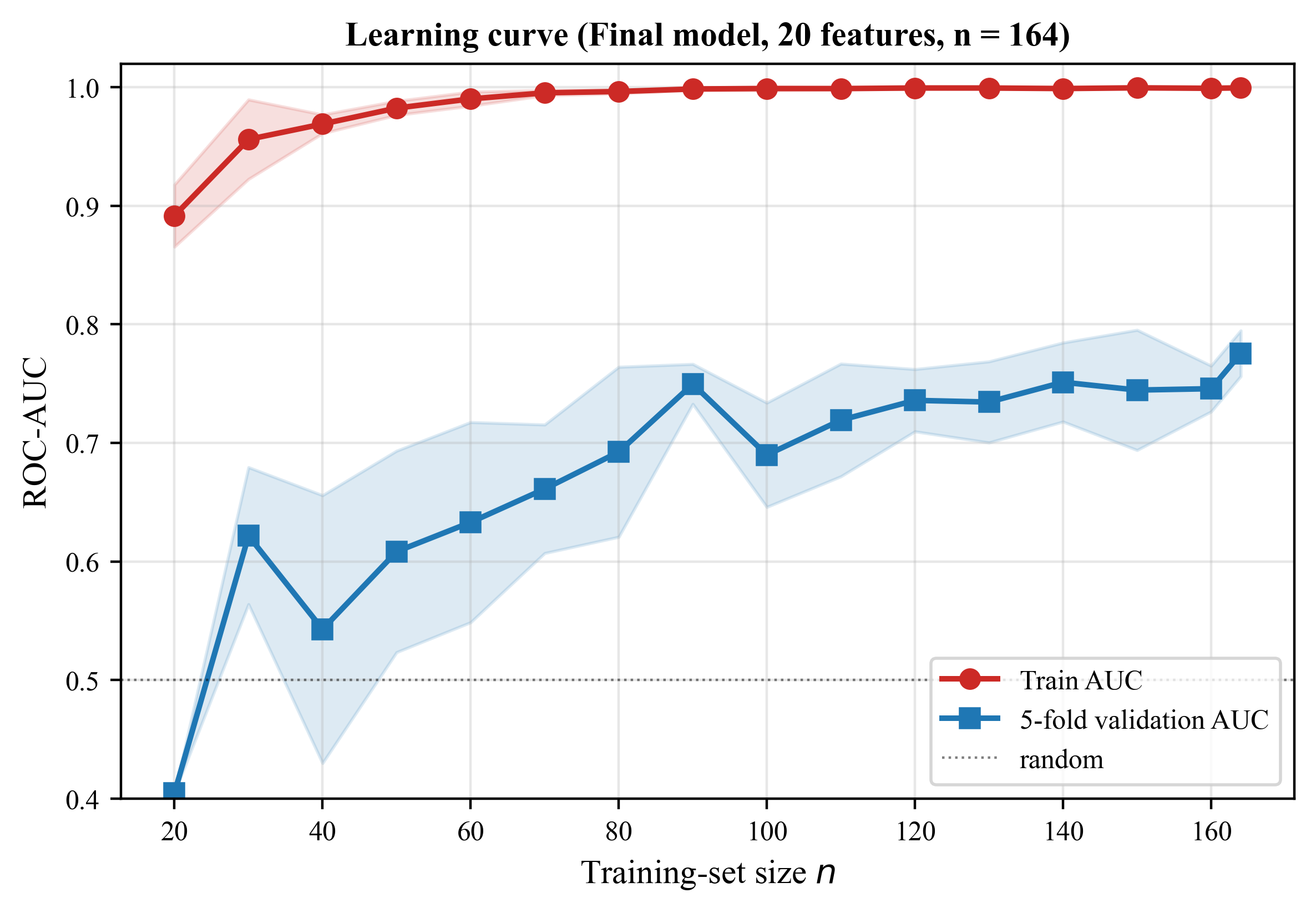}
\caption{Learning curve for the 20-feature Final model under 5-fold stratified CV. The training-set AUC saturates above 0.98 by n $\approx$ 60, while the 5-fold CV AUC continues to climb across the available training-set range, reaching $\approx 0.78$ at n = 164 with a residual train-validation gap of $\approx$ 0.22. The curve has not plateaued: more labelled data is the single highest-leverage intervention available, and this gap suggests that a substantially larger and geographically broader training set would close the bulk of the training-set generalisation deficit.}
\label{fig4}
\end{figure}

Nevertheless, we provide standard 5-fold CV model comparison scores for AUC (Table~\ref{tabII}), ROC and precision-recall curves (Fig.~\ref{fig3}), label-permutation rank-discrimination tests in Section~\ref{sec:results-ablation} (Table~\ref{tabIV}), the pooled 5-fold CV summary in Table~\ref{tabIII} of Appendix~\ref{sec:appx-cv}, and training and validation learning curves in Fig.~\ref{fig4}. We stress, these aggregate model based statistics, and variance based comparisons, are largely uninformative in this setting -- there is considerable overlap in performance between XGboost models (even though they perform better than the L2 logistic model baseline). This is expected, given the extremely small data size used in training: both standard 5-fold CV and label permutation tests are severely underpowered for model comparisons on $\sim$100 data points \cite{ref-braganeto, ref-varoquaux}. As such we are unable to determine that the 20-feature model performs any better than the 2-feature baseline on the training set. Ablation studies in Table~\ref{tabIV} confirm this: engineered POI text features do not improve training-set rank-ordering beyond what the open-data baseline already achieves, and the 2-feature socio-demographic baseline (\texttt{pop\_density} + \texttt{wealth\_rwi}) is essentially tied with the 20-feature Final on the training-set 5-fold CV AUC. Per-fold standard deviations are uniformly large, so differences across models are smaller than within-configuration fold-to-fold variability.

At the same time, we note the ability of the more complex model including POI features to be grounded in more mechanistic explanations than the baseline, its visual correctness in separating predictions based on the underlying features on a 2D surface, its ability to improve prediction accuracy on subsets of the training alongside reductions in variance, and its superior behavior in predicting hold out test data. In the following sections, we explore these additional findings in more detail, to understand how models containing additional POI contextual features are behaving under the hood, and why this might help these models perform better out of sample in completely different regions from which they are trained.

\begin{table}[!t]
\caption{TABLE II. FIVE-FOLD STRATIFIED CV ON THE n = 164 TRAINING SET (XGBoost configurations above the rule, logistic-regression baselines below)}
\label{tabII}
\centering
\setlength{\tabcolsep}{3pt}
\footnotesize
\begin{tabular}{@{}p{2.6cm}rrr@{}}
\hline
\textbf{Configuration} & \textbf{\#feat} & \textbf{AUC} & \textbf{AP} \\
\hline
Open-data baseline & 16 & 0.722 $\pm$ 0.129 & 0.562 $\pm$ 0.188 \\
Full feature bank & 70 & 0.760 $\pm$ 0.114 & 0.546 $\pm$ 0.163 \\
Final model (20 feat.) & 20 & 0.771 $\pm$ 0.083 & 0.569 $\pm$ 0.151 \\
2-feature socio-demographic & 2 & 0.766 $\pm$ 0.077 & 0.672 $\pm$ 0.098 \\
LR --- Final panel & 20 & 0.727 $\pm$ 0.104 & 0.586 $\pm$ 0.133 \\
LR --- open-data baseline & 16 & 0.670 $\pm$ 0.094 & 0.518 $\pm$ 0.043 \\
LR --- 2-feature (pop\_density + wealth\_rwi) & 2 & 0.721 $\pm$ 0.087 & 0.594 $\pm$ 0.103 \\
\hline
\end{tabular}
\end{table}

A brief note on metric choice: AUC and AP are reported rather than recall-at-threshold because the deployment base rate for informal ULAB sites is below $10^{-3}$ per candidate grid cell at the half-kilometre cell sizes used for deployment (Section~\ref{sec:hard-imbalance}). At that prevalence, even high absolute recall and precision produce candidate lists dominated by false positives, so the right deployment metric is precision-at-K with K determined by field-team budget; we return to this when discussing the cross-country test in Section~\ref{sec:results-transfer}.

\subsubsection{Channel-group ablation}
\label{sec:results-ablation}

Table~\ref{tabIV} decomposes the full feature bank by channel family, reporting pooled 5-fold CV AUC, AP, and F1 under the same split. Two readings matter for the broader argument. First, removing the Proprietary-POI channel costs only fold-noise on the training-set 5-fold CV AUC; the engineered text features do not improve training-set rank-ordering beyond what the open-data baseline already achieves. Second, the 2-feature socio-demographic baseline (pop\_density + wealth\_rwi) is essentially tied with the 20-feature Final on the training-set 5-fold CV AUC. The engineered features therefore do not improve the rank-ordering of training samples beyond a simple socio-demographic prior; their value emerges in deployment calibration (Section~\ref{sec:results-ncr}), out-of-region transfer (Section~\ref{sec:results-nbhd}), and spatial selectivity (Appendix~\ref{sec:appx-coverage}, Moran's I). Per-fold standard deviations remain uniformly large, so differences across configurations are smaller than within-configuration fold-to-fold variability and the table should be read for direction rather than significance.

\begin{table}[!t]
\caption{TABLE III. FEATURE-GROUP ABLATION (pooled 5-fold CV, n = 164)}
\label{tabIV}
\centering
\setlength{\tabcolsep}{2.5pt}
\scriptsize
\begin{tabular}{@{}p{2.6cm}rrrr@{}}
\hline
\textbf{Variant} & \textbf{\#feat} & \textbf{AUC} & \textbf{AP} & \textbf{F1} \\
\hline
Full (Overture + Google) & 70 & 0.749 & 0.466 & 0.505 \\
$-$ raw Overture category counts & 65 & 0.746 & 0.452 & 0.510 \\
$-$ socio-demographic & 68 & 0.689 & 0.412 & 0.412 \\
$-$ keyword heuristics & 68 & 0.750 & 0.461 & 0.563 \\
Overture only (open-data baseline) & 16 & 0.718 & 0.483 & 0.545 \\
Google only & 50 & 0.656 & 0.369 & 0.442 \\
Socio-demographic only & 2 & 0.759 & 0.625 & 0.545 \\
Google ULAB score + diversity only & 2 & 0.574 & 0.331 & 0.452 \\
\hline
\end{tabular}
\end{table}

\subsection{Feature importance}
\label{sec:results-fi}

We use the channel taxonomy of Section~\ref{sec:methods-features} (socio-demographic, Open POI, Proprietary POI) and report six complementary lenses: (i) per-positive SHAP attribution at the 50 training positives; (ii) the gap between training-time XGBoost gain and post-hoc SHAP rankings; (iii) a nested ablation across socio-only / $+$Open POI / Final models on the same training set; (iv) the same three lenses applied to 172 held-out TSIP-informal sites that never entered training; (v) the symmetric view at the 114 training negatives; and (vi) a sub-pixel-resolution comparison with the 2.4\,km Meta RWI prior. The first three lenses summarise how the model uses each channel on the training set; (iv) and (v) test whether the same channels remain informative on populations the model never saw; (vi) asks at what spatial scale the engineered features add information.

\subsubsection{Per-positive attribution at the training set}

\begin{figure*}[!t]
\centering
\includegraphics[width=\textwidth,keepaspectratio]{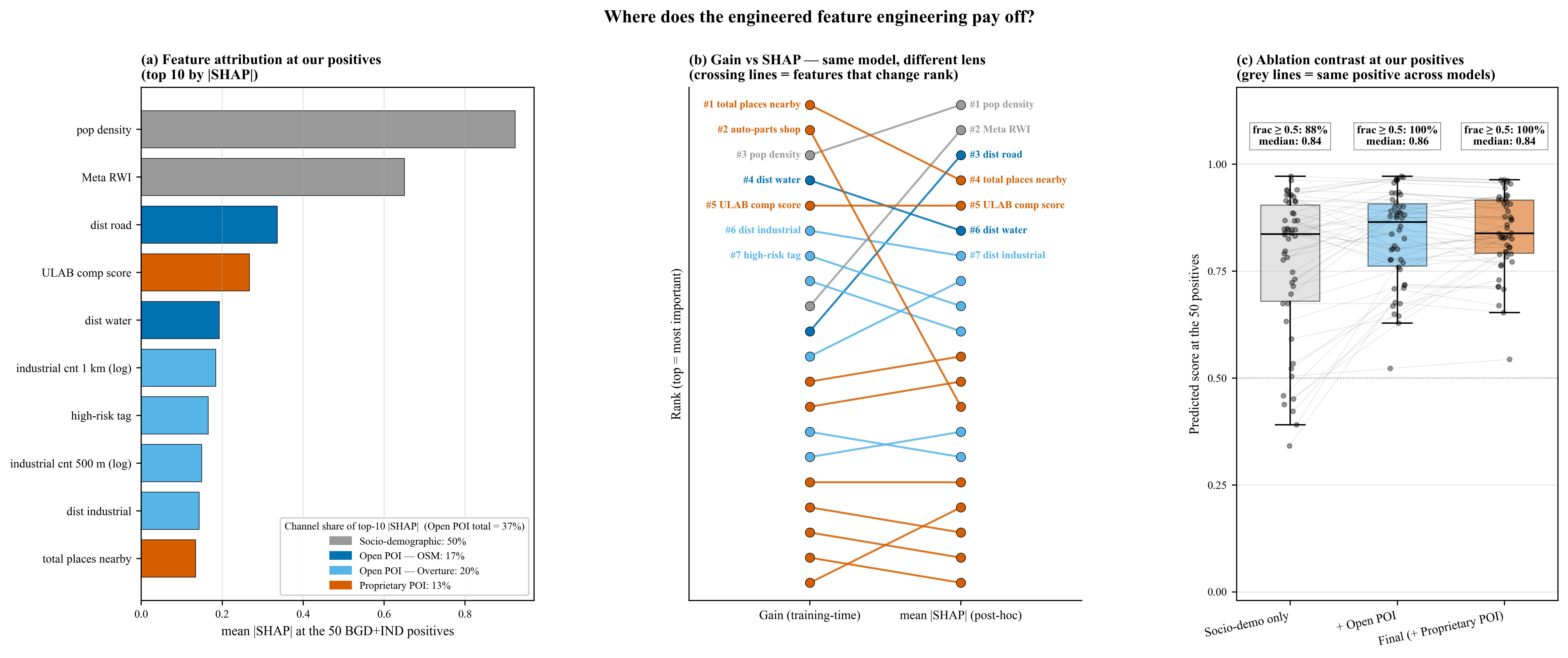}
\caption{Three lenses on feature importance for the Final model (n = 164: 33 BGD + 17 IND positives, 114 negatives). \textbf{(a)} Top-10 features by mean $|\mathrm{SHAP}|$ at the 50 BGD+IND training positives, coloured by channel family (socio-demographic in grey, Open POI in two blue shades for OSM proximity vs Overture, Proprietary POI in vermillion); at the training positives, socio-demographic priors are the dominant explainer. \textbf{(b)} Gain rank (training-time, left) vs.\ mean $|\mathrm{SHAP}|$ rank (post-hoc, right). The Proprietary-POI feature \texttt{goog\_v2\_total\_places} tops the gain ranking but ranks \#4 on SHAP, while \texttt{pop\_density} climbs from gain rank \#3 to SHAP rank \#1: the Proprietary-POI channel does its work near the root of each tree (high gain), and downstream socio-demographic splits absorb part of its partitioning work. \textbf{(c)} Predicted score on the 50 training positives under three nested models. The fraction at or above $\tau = 0.5$ rises from 88\,\% under socio-demographic features alone to 100\,\% after adding Open POI; the Proprietary-POI channel does not add further threshold rescue on this set.}
\label{fig9}
\end{figure*}

TreeSHAP \cite{ref28} attributions at the 50 BGD+IND training positives (Fig.~\ref{fig9}(a)) place roughly half the top-10 attribution mass on the 2 socio-demographic features, about a third on Open POI, and the remainder on Proprietary POI. The Section~\ref{sec:methods-features} caveat applies: WorldPop and Meta RWI are themselves machine-learning products built in part from urban-commercial signals (cellphone handsets, road density, building footprints) \cite{ref5, ref6}, so part of what reads as ``socio-demographic'' attribution silently embeds POI-adjacent information. 

\subsubsection{Gain vs SHAP: training-time splitter vs marginal contribution}

\begin{table}[!t]
\caption{TABLE IV. TOP-10 FEATURES BY XGBOOST GAIN (Final model, n = 164)}
\label{tabVII}
\centering
\resizebox{\columnwidth}{!}{%
\begin{tabular}{cllr}
\hline
\textbf{Rank} & \textbf{Feature} & \textbf{Source} & \textbf{Gain} \\
\hline
1 & goog\_v2\_total\_places & Google Places & 13.39 \\
2 & goog\_v2\_pt\_auto\_parts\_store & Google Places & 3.85 \\
3 & pop\_density & WorldPop & 3.46 \\
4 & dist\_water\_m & OSM & 3.07 \\
5 & goog\_v2\_ulab\_score & Google Places & 3.03 \\
6 & dist\_industrial\_m & OSM & 2.91 \\
7 & is\_high\_risk\_tag & Overture & 2.90 \\
8 & industrial\_count\_500m\_log & Overture & 2.82 \\
9 & wealth\_rwi & WorldPop/RWI & 2.57 \\
10 & dist\_road\_m & OSM & 2.25 \\
\hline
\end{tabular}}
\end{table}

Comparing XGBoost Gain (Table~\ref{tabVII}) with mean |SHAP| (Fig.~\ref{fig9}(b)) exposes a distinct behaviour of POI features. While the hyper-local \texttt{goog\_v2\_total\_places} ranks first in Gain, it drops to fourth in SHAP. This divergence stems from how the model handles scale: Gain rewards the high-granularity proprietary feature for its capacity to isolate localized variance and drive deep training loss reduction early in tree construction. Conversely, SHAP evaluates marginal predictive contributions across all feature coalitions. Because the coarser socio-demographic features establish the baseline regional gradients, SHAP de-emphasizes the granular POI data, attributing the bulk of the predictive lift to the macro-level indicators. Ultimately, \texttt{goog\_v2\_total\_places} still functions as a critical structural scaffold—providing the fine-grained partitioning likely required for downstream precision on test data (see below), even if its marginal impact is eclipsed globally by coarser spatial predictors.

\subsubsection{Nested ablation: who clears the threshold}

A nested-model ablation on the same training set isolates each channel's contribution to the score directly (Fig.~\ref{fig9}(c)). Going from a 2-feature socio-demographic baseline to $+$Open POI moves the lower tail of the predicted-score distribution above $\tau = 0.5$, taking the count of positives at or above threshold from 44/50 to 50/50. Whilst this seems small, the distributional tightening with the additional POI data is notable. Adding the Proprietary-POI channel on top does not change this count further. The reading combines with Table~\ref{tabIV} (Section~\ref{sec:results-ablation}): the Proprietary-POI channel costs only fold-noise on training AUC when removed, so its training-set contribution is not in threshold rescue here but in early-tree splitting (the gain view above) and in out-of-region selectivity (next subsubsection).

\subsubsection{Held-out TSIP transfer: the Proprietary-POI channel matters more on the held-out testing populations}

Applying the same three lenses to 172 held-out TSIP-informal sites (36 in non-NCR India, 136 in Bangladesh, none in training) offers further insight. Relative to the training set, the Proprietary-POI share of mean $|\mathrm{SHAP}|$ attribution rises at every out-of-training population, peaking at the 136-site Bangladesh subset: where Overture and OSM coverage is likely sparser (Section~\ref{sec:hard-geo}, Fig.~\ref{fig1}), the model leans more heavily on Google Places features to compensate. The cross-population summary across all six populations measured in this paper is in Fig.~\ref{fig15} (Appendix~\ref{sec:appx-summary}). The aggregate fraction at or above $\tau = 0.5$ on the 172 held-out positives stays flat across the three nested models, but the Proprietary-POI channel pulls a handful of sites (most of them in Bangladesh) across the threshold that the Open-POI-only model misses, balanced by a similar number of false negatives elsewhere. These wins are interesting.

\subsubsection{Symmetric view at the training negatives}

\begin{figure*}[!t]
\centering
\includegraphics[width=\textwidth,keepaspectratio]{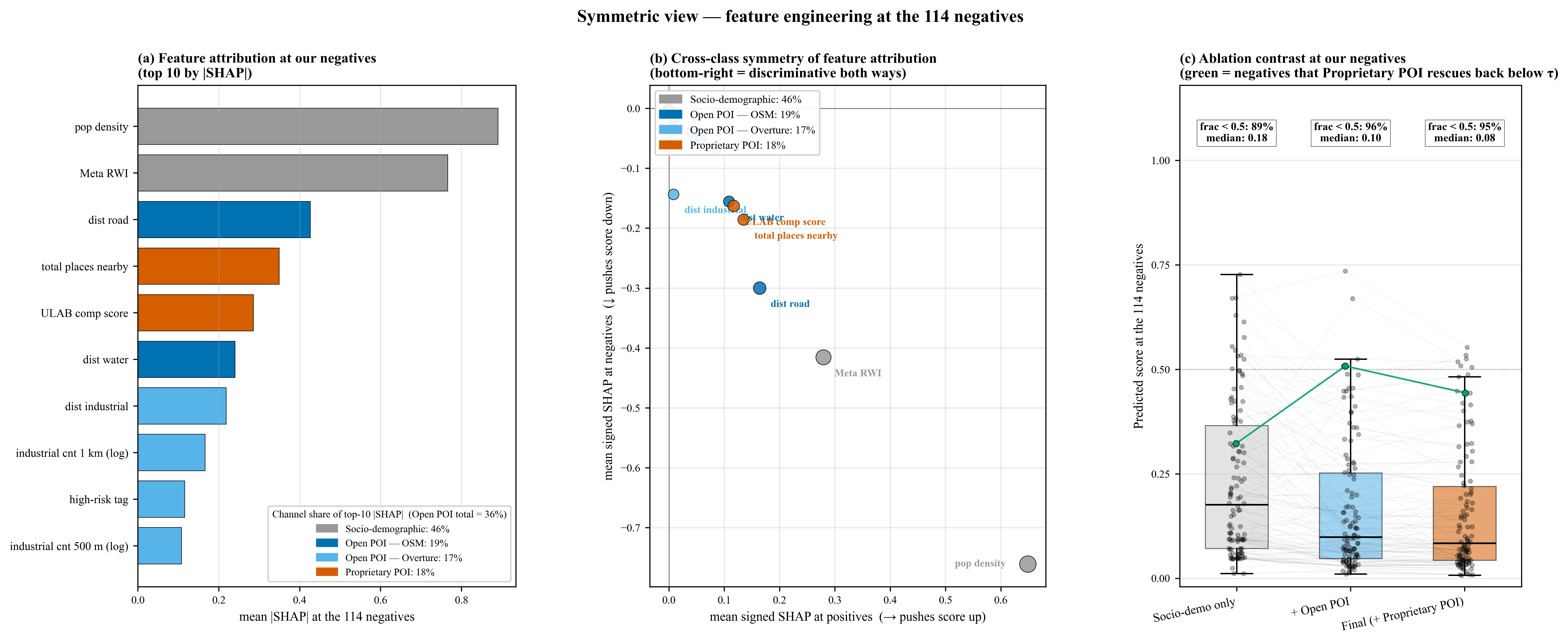}
\caption{Symmetric view of feature importance at the 114 training negatives. \textbf{(a)} Top-10 features by mean $|\mathrm{SHAP}|$, channel-coloured; the Proprietary-POI share is modestly higher than at positives, consistent with the absence of POI counts being a real ``not-ULAB'' signal as well as the presence of POI counts being a ``ULAB-like'' one. \textbf{(b)} Cross-class signed-SHAP scatter (positives on the x-axis, negatives on the y-axis); the bottom-right quadrant identifies features discriminative in both directions, where \texttt{pop\_density} sits decisively. \textbf{(c)} Nested ablation at the 114 negatives: the fraction below $\tau = 0.5$ rises from 89\,\% (socio only) to 96\,\% ($+$Open POI) and sits at 95\,\% under the Final model. The Proprietary-POI channel pulls one negative back below threshold and introduces one above elsewhere; the net effect at $\tau = 0.5$ is approximately neutral, but the negative-side median tightens.}
\label{fig11}
\end{figure*}

The symmetric question of which features keep negatives below the threshold yields a picture that broadly parallels the positive-side analysis (Fig.~\ref{fig11}). The channel-share at the negatives is more evenly distributed than at positives, with a modestly higher Proprietary-POI share: the absence of POI counts is itself a ``not-ULAB'' signal. The cross-class signed-SHAP scatter (Fig.~\ref{fig11}(b)) places \texttt{pop\_density} in the bottom-right quadrant (pushes positives up, negatives down), while Open POI and Proprietary-POI features cluster near the negative-pushing axis. The nested ablation at the negatives (Fig.~\ref{fig11}(c)) tracks the positive-side trajectory: Open POI does most of the work, the Proprietary-POI channel adds score-distribution concentration rather than threshold rescue. Both POI based models tighten model predictions for negatives as they did for positives.

\subsubsection{Resolution: sub-pixel discrimination beyond the wealth prior}

\begin{figure*}[!t]
\centering
\includegraphics[width=\textwidth,keepaspectratio]{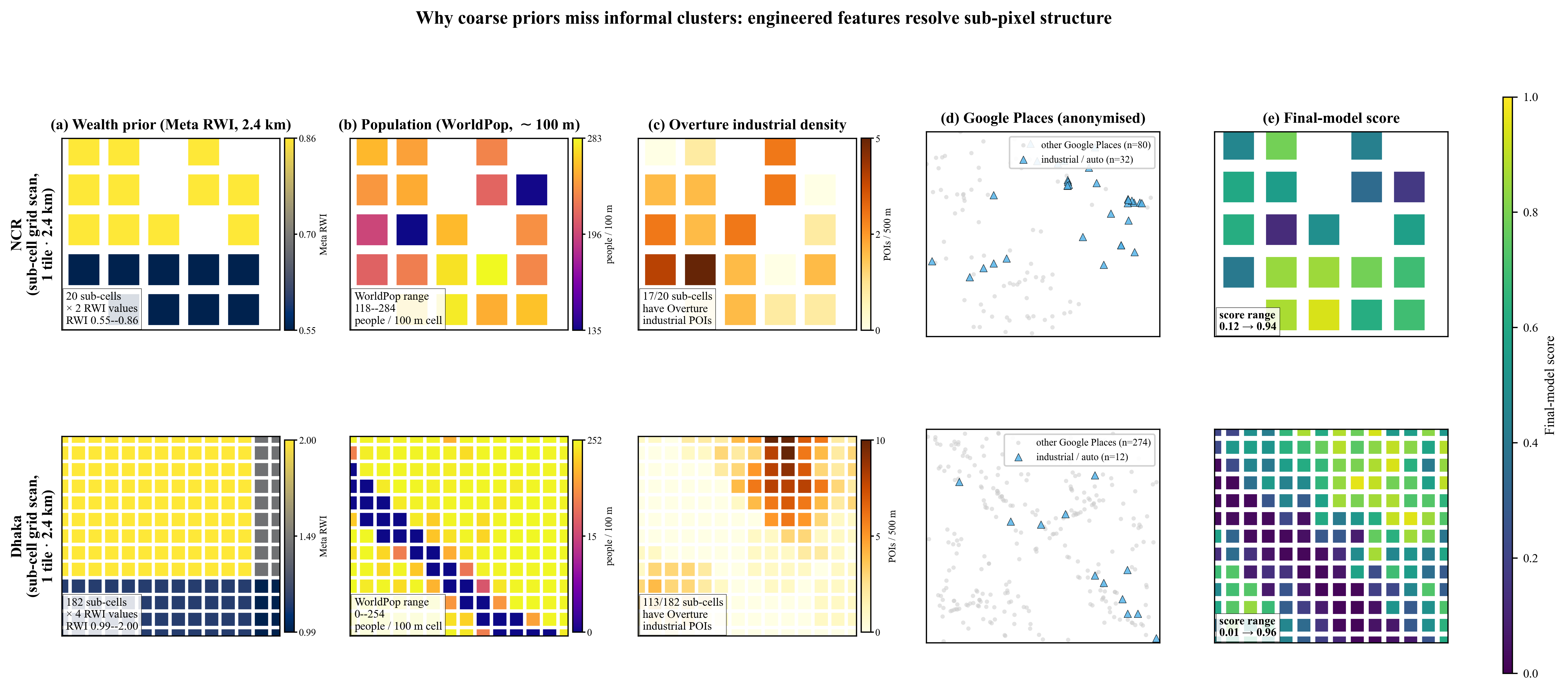}
\caption{Why coarse priors miss informal clusters. The scan grid is at 200\,m sub-cell spacing in every panel. NCR Sahibabad on top (20 sub-cells) and Old Dhaka on the bottom (182 sub-cells). Panels (a)--(e) show, in order: the wealth prior (Meta RWI, ~2.4\,km native pixel), the population layer (WorldPop, ~100\,m native pixel), the Overture industrial POI-density (count of Overture POIs matching the industrial keyword union within a 500\,m buffer), anonymised google Places points, and the Final-model score grid. Note: the industrial-density panel here uses Overture POIs, not OSM.}
\label{fig10}
\end{figure*}

A complementary spatial view (Fig.~\ref{fig10}) scans two single 2.4\,km Meta RWI pixels at 200\,m sub-cell spacing (one in NCR Sahibabad, one in Old Dhaka). Within each tile, the wealth prior are a couple of adjacent native pixels, but the Final-model score spans roughly an order of magnitude. The two rows give complementary diagnostics of what the engineered POI features are doing under the hood. In the NCR row the population layer and the score field carry visibly related structure, but the score also resolves sub-zone variation that population alone does not, the Overture industrial-density panel and the Google Places panel are doing additional partitioning work. In the Dhaka row the contrast is much sharper: the upper half of the tile is uniformly higher-wealth and higher population density (yellow in panels a and b), but the Final-model score picks up specific sub-cells that track the industrial clusters from POIs, the engineered features add discriminative information at finer spatial scales than either socio-demographic surface expresses. This is consistent with the deploymentcalibration tightening reported in Section~\ref{sec:results-ncr}.

\subsection{Cross-country transfer}
\label{sec:results-transfer}

Two complementary cross-country tests are evaluated. The standard leave-one-country-out (BGD$\to$IND and IND$\to$BGD) yields AUCs near chance (0.61--0.67; details in Appendix~\ref{sec:appx-cv}), indicating that some of the training-set signal is country-specific. With only two source countries this is more of a curiosity than anything to determine deployment readiness.

\begin{figure*}[!t]
\centering
\includegraphics[width=\textwidth,keepaspectratio]{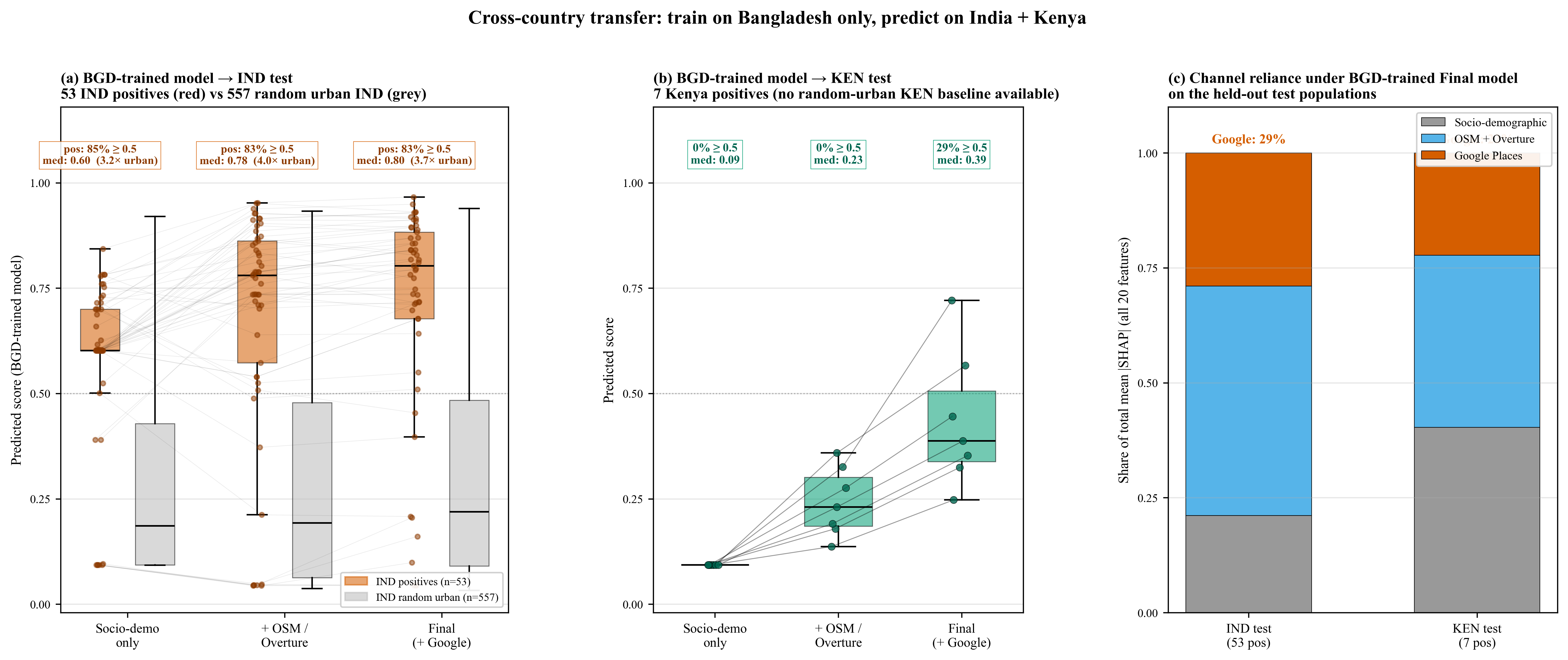}
\caption{Cross-country transfer: three nested models trained on Bangladesh-only data ($n = 78$), predicting on out-of-country positives. \textbf{(a)} 53 held-out India positives (red) against 557 IND random-urban controls (grey); all three models separate positives from controls highly significantly. \textbf{(b)} 7 Kenya positives; only the Final model lifts any (2/7) above $\tau = 0.5$. \textbf{(c)} Channel share of mean $|\mathrm{SHAP}|$ at the held-out IND and KEN test positives. At the IND test the Proprietary-POI share reaches 29\,\% (the highest in the paper) with Open POI taking 50\,\%; at KEN the share tilts toward socio-demographic.}
\label{fig13}
\end{figure*}

That said, for deployment, a relevant variant was explored that retrains the three nested models on Bangladesh-only data (33 positives, 45 negatives) and predicts on every Indian and Kenyan TSIP-listed positive for which features are available (Fig.~\ref{fig13}). On the 53 held-out IND positives, all three nested models produce strong separation from random-urban IND controls: median score lifts of $\approx$ 3--4$\times$, with one-sided Mann--Whitney $p < 10^{-15}$ under the independence assumption (reported as exploratory; the same spatial-autocorrelation caveat as in Section~\ref{sec:results-nbhd} applies). Most of this performance in IND comes from socio-demographic and Open POI features, with the Proprietary-POI channel adding only a small further median lift. On the 7 Kenya positives the picture inverts: the socio-only and $+$Open POI models leave every site below $\tau = 0.5$, and only the Final model rescues 2 of the 7. With $n = 7$ the counts are too small to support a quantitative claim, but this is the only test in the paper where the Proprietary-POI channel is necessary to move any positive across the threshold. 

The IND-test trajectory is also bidirectional. As channels are added the median rises, but the headline fraction at $\tau = 0.5$ is essentially flat. Some of these flips are plausibly informative (e.g., a high-population urban centre without supporting POI infrastructure is not where informal recycling clusters), while others are likely artefacts of POI-coverage bias or hazard-specific signals that current lexicon might not yet encode; this is exactly the kind of failure mode the open-problem roadmap of Section~\ref{sec:roadmap} is designed to interrogate. A threshold-tuned deployment would also recover most of the just-below-threshold sites: at the field-team-budget-dependent precision-at-K regime introduced in Section~\ref{sec:results-cv} the operator can lower $\tau$ to widen the candidate list. Channel-share of $|\mathrm{SHAP}|$ at the cross-country test populations (panel c; cross-population summary in Fig.~\ref{fig15}, Appendix~\ref{sec:appx-summary}) shows the Proprietary-POI share rising at every out-of-training population relative to the training set, consistent with the engineered POI channels carrying a larger share of attribution the further the prediction sits from the training distribution. Increasing training-country diversity is one of the open problems set out in Section~\ref{sec:roadmap}.

\subsection{NCR test data deployment}
\label{sec:results-ncr}

To test geographic transfer, each trained model was applied without retraining to India's National Capital Region. A regular grid at $0.0045^\circ$ spacing ($\approx$500\,m) was laid across a bounding box from $(76.60^\circ\mathrm{E},\,28.40^\circ\mathrm{N})$ to $(77.75^\circ\mathrm{E},\,29.47^\circ\mathrm{N})$ (60{,}435 cells), then filtered to 3{,}746 cells ($\approx$6\,\%) retained after a population-density and OSM-presence screen (WorldPop > 0 inhabitants and at least one OSM road or POI within 1\,km). The filter concentrates the scan on the populated NCR area; features were extracted identically to the training pipeline. Ground truth consists of 10 GPS-precise TSIP-georeferenced ULAB sites \cite{ref7}, none of which were in training.

Table~\ref{tabV} reports the resulting score distributions. The open-data baseline flags about half the grid above the deployment threshold; adding the Google Places multi-channel text features tightens this distribution by roughly $5\times$; the Final model sits between the two. The deployment-region negatives carry most of the calibration improvement (without them, the open-data baseline flags $\approx$ 90\% of NCR cells); the Google Places text features then sharpen the within-deployment ranking. 

Compared across deployment-regime diagnostics (NCR fraction-above-threshold, neighbourhood-validation cell counts at TSIP sites, and Moran's I on the score field; see Appendix~\ref{sec:appx-coverage}, Table~\ref{tabMoran}), the simpler open-data baseline produces noticeably more over-confident and less spatially structured deployment behaviour -- even though it appeared to perform just as well using training evaluations of AUC/ROC and rank-permutation tests. The engineered features therefore can clearly add value the training AUC contrast does not see, in selectivity and in spatial coherence, and that distinction is one of the empirical observations the paper rests on (Sections~\ref{sec:results-fi}, \ref{sec:results-transfer}, \ref{sec:results-nbhd}).

\begin{table}[!t]
\caption{TABLE V. NCR DEPLOYMENT CALIBRATION (fraction of 3,746 grid cells above threshold)}
\label{tabV}
\centering
\setlength{\tabcolsep}{3pt}
\footnotesize
\begin{tabular}{@{}lrrr@{}}
\hline
\textbf{Threshold} & \textbf{Open-data} & \textbf{Full bank} & \textbf{Final model} \\
\hline
$\geq$ 0.3 & 0.709 & 0.125 & 0.518 \\
$\geq$ 0.5 & 0.501 & 0.091 & 0.289 \\
$\geq$ 0.7 & 0.287 & 0.051 & 0.123 \\
$\geq$ 0.9 & 0.051 & 0.009 & 0.018 \\
\hline
\end{tabular}
\end{table}

\subsection{Neighbourhood validation against TSIP ground truth}
\label{sec:results-nbhd}

This subsection presents three independent external-validation tiers. The first is an area-based test against the 10 TSIP-georeferenced informal sites inside the NCR deployment region. The second is an out-of-region test on 172 TSIP-documented informal sites in non-NCR India and Bangladesh that never entered training. The third is a fully out-of-source test against 131 regulatory-registered formal recyclers scraped from Indian state-level registries. The methodology produces a measurable signal in all three; the strongest are the non-NCR India out-of-region test (TSIP-informal sites flagged at roughly $5\times$ the random-control rate) and the formal-vs-informal discrimination (informal sites scoring $2.3\times$ formal recyclers in non-NCR India). We report Mann--Whitney $p$-values alongside these effect sizes for reference, but read them as exploratory indicators only: per-cell statistics share strong spatial autocorrelation (Appendix~\ref{sec:appx-coverage}, Table~\ref{tabMoran}) and the comparator sample sizes are modest, so the independence assumption underlying the $p$-values is violated.

Informal ULAB recyclers operate at sub-grid-cell scale (a single courtyard or workshop, 5--50 m), but their economic ecosystem (auto-repair shops, scrap markets, battery resellers) extends across 1--5 km clusters. The contextual features summarise this surrounding environment, not the recycling unit. Per-point recall would conflate ground-truth GPS uncertainty with model spatial resolution. TSIP is also a sparse, single-source sample of confirmed positives; the absence of TSIP coverage in a sub-area is not evidence of absence. The right metric is therefore enrichment of high-score cells in TSIP-documented sub-zones relative to random NCR neighbourhoods, not specificity against an assumed-comprehensive negative set. For each ground-truth location $x$ and buffer radius $R \in \{1, 2, 5\}$\,km we compute the maximum, mean, and high-confidence count $|\{j : s_j \geq 0.7\}|$ of Final-model scores over the haversine ball $B(x, R)$ and compare against 10{,}000 random NCR neighbourhoods; one-sided empirical $p$-values are the fraction of null neighbourhoods meeting or exceeding the TSIP-site mean.

The numerical breakdown is reported in Table~\ref{tabNbhd} of Appendix~\ref{sec:appx-nbhd}, with the NCR risk surface and null-vs-observed distributions shown in Figs.~\ref{fig5} and~\ref{fig6}. On the 10 TSIP-georeferenced NCR sites the area-based test exhibits a regional rather than a site-level pattern. Lead-effect-size: at the 5\,km buffer the maximum Final-model score within the TSIP neighbourhood is $1.12\times$ the random-NCR null mean (0.93 vs 0.83); at the 2\,km buffer it is $1.18\times$ (0.78 vs 0.66); the high-score-cell-count metric does not separate at any buffer. Read together with the spatial-coherence analysis (Appendix~\ref{sec:appx-coverage}, Moran's I in Table~\ref{tabMoran}), the model reliably finds at least one high-confidence cell in the broader sub-region around each documented site, even if the count of such cells is not distinguishable from the null. The relative separation is modest ($\sim$10--20\% above the random baseline) and the comparison is severely underpowered ($n = 10$ TSIP-NCR sites) with strong spatial autocorrelation between neighbouring cells, so the per-buffer $p$-values in Appendix~\ref{sec:appx-nbhd} should be read as exploratory ranking indicators rather than confirmatory tests; the substantive observation is the regional-scale enrichment, not the precise significance level.

The out-of-region TSIP global test is the strongest external-transfer signal in the paper. The complete 20-feature panel was extracted, with no retraining, for 36 TSIP-georeferenced ULAB sites in non-NCR India, 136 informal Bangladesh TSIP sites, and 1{,}157 pop\_density-matched random urban controls split across both countries. Two tests are reported per country: TSIP-score distribution against random urban controls, and against population-density-matched controls (within $\pm$20\% population quantile of each TSIP site). The India non-NCR result is unambiguous in effect-size terms: TSIP sites are flagged at roughly $5\times$ the rate of random controls in both tests (one-sided Mann--Whitney $p < 0.0001$, reported as exploratory). Bangladesh transfer is present but materially weaker: separation is on the order of $1.5\times$ above the matched-control baseline ($p \approx 0.03$), reflecting a hard-negative-mining trade-off in which BGD TSIP-formal hard negatives partly overlap BGD informal positives in feature space, so adding too many of them weakens transfer. The headline configuration sacrifices some training-set sharpness to recover Bangladesh transfer. As above, the reported $p$-values assume independent samples; spatial autocorrelation within and between scan regions (Appendix~\ref{sec:appx-coverage}) makes them exploratory rather than confirmatory.

\begin{figure}[!t]
\centering
\includegraphics[width=\columnwidth,keepaspectratio]{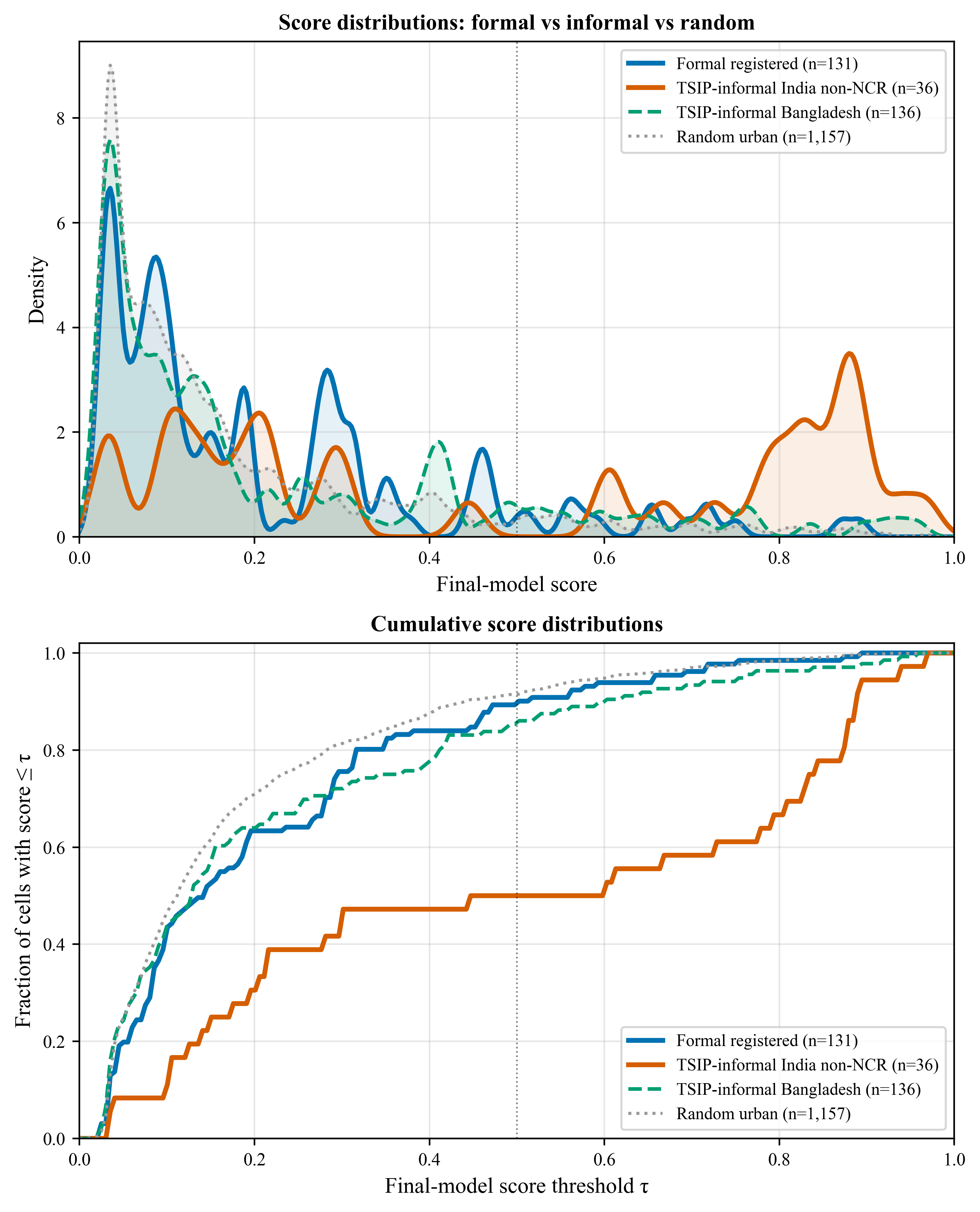}
\caption{Score distributions of the Final model on four independent populations: 131 confirmed-FORMAL recyclers from Indian regulatory registries (CPCB / MPCB / DPCC / TNPCB), 36 TSIP-informal sites in non-NCR India, 136 TSIP-informal sites in Bangladesh, and 1,157 random urban controls in BGD + IND. Left: kernel density estimates. Right: cumulative form (fraction with score $\leq$ $\tau$). The TSIP-informal non-NCR India distribution exhibits a clear secondary mode above 0.7 absent from all other groups; formal registered facilities sit at approximately the random-urban baseline. The Bangladesh informal distribution sits closer to formal/random, reflecting the BGD out-of-region regression.}
\label{fig8}
\end{figure}

To address the TSIP source-bias concern (Section~\ref{sec:results-threats}), a fully independent negative validation set was constructed by scraping four official Indian recycler registries (CPCB, MPCB, DPCC, TNPCB). After geocoding, deduplication, and exclusion of training overlaps, 131 confirmed-formal lead-acid-battery recyclers were extracted with the full 20-feature panel: 31 within the NCR scan bbox and 100 stratified across other Indian states. None entered any training, validation, or hard-negative pool. Table~\ref{tabFormal} reports three formal-vs-informal hypotheses tested on this set; score distributions are shown in Fig.~\ref{fig8}. We report relative mean separation (Sep.\ column) as the primary effect-size indicator, with one-sided Mann--Whitney $p$-values alongside for reference. The two strongest findings: in the NCR deployment region, formal-registered facilities score $1.54{\times}$ lower than the average NCR cell, so the model is not simply flagging industrial-like urban activity. In non-NCR India, TSIP-informal sites score $2.28{\times}$ higher than formal recyclers at approximately the random-urban baseline. The combined hypothesis (formal vs all TSIP-informal) shows only modest $1.39{\times}$ separation, driven by the known Bangladesh weakness where informal and formal scores both sit near the random baseline. We caveat the reported $p$-values: neighbouring urban cells share POI context and the score field is spatially autocorrelated (Moran's I in Table~\ref{tabMoran}), so Mann--Whitney's independence assumption is violated and we read the $p$-values as exploratory hypothesis indicators rather than confirmatory tests. The separation ratios are reported as the primary takeaway: this out-of-source external validation is the strongest negative-side evidence in the paper that the methodology captures informal-recycler-specific structure rather than generic industrial signal.

\begin{table*}[!t]
\caption{TABLE VI. FORMAL-VS-INFORMAL DISCRIMINATION TESTS (Final model, 131 registry-confirmed formal sites vs informal / random comparators)}
\label{tabFormal}
\centering
\resizebox{\textwidth}{!}{%
\begin{tabular}{p{4.4cm}p{5.5cm}cccc}
\hline
\textbf{Hypothesis} & \textbf{Comparator} & \textbf{Mean (formal)} & \textbf{Mean (comparator)} & \textbf{Sep.} & \textbf{MW $p$} \\
\hline
H1: NCR formal vs NCR grid & 31 NCR formal vs 3{,}746 NCR cells & 0.205 (10\% $\geq$ 0.5) & 0.315 (23\%) & 1.54$\times$ & 0.0007 \\
H2: Non-NCR formal vs non-NCR informal & 100 IND non-NCR formal vs 36 TSIP-informal IND & 0.217 & 0.495 & 2.28$\times$ & $<$ 0.0001 \\
H3: Formal vs all TSIP-informal combined & 131 formal vs 172 TSIP-informal (BGD+IND) & 0.214 & 0.298 & 1.39$\times$ & 0.123 \\
\hline
\end{tabular}}
\end{table*}

To understand which channels drive these results, we re-ran the same five validation hypotheses (H2--H6, listed in Table~\ref{tabNested} of Appendix~\ref{sec:appx-summary}; Fig.~\ref{fig15} panel (b) gives a heatmap view) under each of the three nested models. The qualitative pattern is consistent across tests: the engineered POI channels increase the relative separation between informal-positive and comparator populations in four of the five tests, with the most pronounced effect on the Bangladesh out-of-region test (H5), which shows essentially no separation under socio-demographic features alone but recovers detectable separation once the Proprietary-POI channel is added. The Indian non-NCR tests (H2, H4) and the NCR 5\,km neighbourhood test (H6) all show monotone separation gains from socio-only to Open-POI to Final. The combined formal-vs-informal test (H3) is the exception: separation remains modest under every model because the Bangladesh informal sub-population sits near the formal and random-urban baselines and dilutes the combined effect size. As above, neighbouring urban cells share POI context and the score field exhibits spatial autocorrelation (Table~\ref{tabMoran}), so the per-cell Mann--Whitney $p$-values in Table~\ref{tabNested} should be read as exploratory ranking indicators of channel contribution rather than as confirmatory tests; the takeaway is the direction and magnitude of the change in separation as channels are added, not the absolute $p$ values.

\subsection{Threats to validity}
\label{sec:results-threats}

\subsubsection{Data limitations}

The empirical claims in this paper rest on a small, single-sourced label set and on a desk-based evaluation. Both training positives and out-of-region testing positives come from Pure Earth's TSIP database, so any TSIP source-bias likely propagates through every result; the 131-site regulatory-registry formal-recycler validation (Section~\ref{sec:results-nbhd}) addresses this on the negative side but not on the positive side. The 50 training positives are also heterogeneous in scale and operational type (household-courtyard battery-breaking through to semi-formal collection sites), treated as a single class. Google Places coverage and OSM completeness both correlate with economic development, so the deployment-calibration improvement of Section~\ref{sec:results-ncr} and the training-set discrimination above the open-data baseline partly reflect richer POI ecologies in higher-coverage regions; the coverage-tertile analysis in Appendix~\ref{sec:appx-coverage} bounds, but does not remove, this caveat. Cross-region transfer beyond South Asia is empirically untested; bridging it would require country-specific lexica and re-tuned socio-demographic priors. Finally, every empirical claim here is desk-based: no predicted high-confidence cell has been field-verified by XRF. The methodology's failure modes are characterised against existing labels rather than by prospective discovery, and prospective field-verification is one of the open problems in Section~\ref{sec:roadmap}.

\subsubsection{Model assumptions}

Similarly, we have not undertaken sophisticated modeling of correlated error in our data during our demonstrations and explorations (largely itself due to data limitations). This means our results should be interpreted not as final but as indicative of potential hypotheses to explore when more data comes available. For example, our initial tests do suggest that future work building on larger datasets would need to account for correlated error. First, the accuracy statistics we present based on simple stratified CV are likely inflated: positives drawn from the same city share correlated POI context, so blocking by cluster (e.g.\ the Delhi/NCR cohort) exposes the distribution shift \cite{ref26} that tempers zero-shot deployment performance. Further, while spatial-coherence analyses using Moran's I show the model successfully discriminates targeted sub-clusters rather than merely proxying general urbanisation, the spatial structure within the scores suggests a need to account for spatial autocorrelation explicitly (Appendix~\ref{sec:appx-coverage}, Table~\ref{tabMoran}). Finally, we find that the methodology is not driven by coverage differences in features (e.g.\ Google Places) but is bounded by them (full numbers in Appendix~\ref{sec:appx-coverage}) -- indicating a bias structure in the features that could be accounted for in modelling more explicitly.

\section{A Roadmap of Open Problems}
\label{sec:roadmap}

The empirical findings of Section~\ref{sec:results} indicate how contextual geospatial features can help improve identification of informal environmental hazards, showing with a ULAB demonstration where the methodology works (channel attribution, threshold rescue, deployment-region selectivity) and where the pipeline is fragile (small-sample training-set AUC, cross-country transfer beyond South Asia, single-source NCR validation). The stakes are large: lead exposure alone accounts for an estimated 5.5 million adult cardiovascular deaths and on the order of \$6 trillion in annual economic losses, much of it traceable to informal ULAB recycling in LMICs \cite{ref10}. The open problems below pull these threads together: the deployment-relevant signal is built from engineered, expert-validated features rather than from a generic socio-demographic prior, and scaling it requires more domain-expert collaboration at least as much as more raw data.

\subsection{Ground-truth acquisition at scale}

Clearly, limited labels prohibit robust generalisation. Scaling ground truth for informal environmental hazards requires coordination with field organisations (Pure Earth, Toxics Link, UNICEF, national agencies) for standardised GPS-precise annotations; active-learning frameworks \cite{ref20} could prioritise the most informative sites. While we have motivated our demonstration with ULAB, to our knowledge no field inventory comparable in scale to TSIP exists, for e-waste burning, ASGM, brick kilns, or tanneries. Partial datasets do exist via informal-economy NGOs, public-health surveys, and academic field studies, but not at the scale needed for quantitative model fitting across regions. Either bootstrapping from these partial sources or designing a class-wide field-investigation programme analogous to Pure Earth's is the single largest barrier to applying the methodology beyond ULAB. This is critical if this problem is to be solved in the next decade.

\subsection{Develop more robust models}

Spatial cross-validation \cite{ref18, ref19} is essential for robust transfer of inference and confidence. Within this, spatial autocorrelation aware CV should replace standard stratified CV. While blocked spatial CV within a country is an intermediate option until more training countries are available, ideally additional countries are needed for LOOCV to be performed, and for spatial correlation structures to be determined. Even before that, repeated $k$-fold CV (e.g.\ 10 repeats of the same 5-fold split with different random seeds) would partially address the fold-to-fold instability we observe at $n \approx 100$ and is the cheapest near-term step on the same training set. Notably, no public benchmark dataset exists for any below-resolution informal-hazard class. \textit{For the broader class:} the absence of such a benchmark makes cross-method comparison and evaluation of robustness difficult; establishing one, through for example a set of benchmark datasets, would let the broader research community contribute methods comparably.

\subsection{The POI-text bottleneck}
\label{sec:roadmap-poi-text}

We find that text-level POI matching outperforms category counts on cross-country transfer and deployment calibration, marginally on the training-set metric. Language-based semantic matching against POI names performs better than pre-defined categorical schemas. Additional improvements could include multilingual embeddings of POI names with hazard-specific query vectors; systematic expansion of local-language keyword lexica (Swahili, Bahasa, Amharic, Hausa); quantification of Google Places coverage bias across LMIC income brackets. For many informal environmental hazards the POI-text bottleneck is class-wide. Each hazard has its own informative POI ecosystem (electronics dismantlers for e-waste, sluice-and-amalgam vendors for ASGM, kiln operators for brick kilns, hide-and-skin dealers for tanneries) and is subject to the same constraints: local-language lexica, embedding-based semantic search, and adequate business-name coverage.

\subsection{Cross-continent generalisation}

Training data is entirely South Asian. Deployment elsewhere faces two concerns: POI coverage drops sharply outside South Asia \cite{ref14}, and morphological patterns of ULAB recycling may differ. Open questions: can domain-adaptation techniques preserve signal without target-country labels? What is the minimum labelled-data budget per new geography? Can multimodal foundation models such as AlphaEarth \cite{ref29} provide geography-invariant representations that contextual features alone cannot? Can sub-Saharan Africa, Latin America, and Southeast Asia models retrained as a multi-region classifier rather than a South-Asia-specific one. For all informal environmental hazards: whether geography-invariant representations exist for any below-resolution informal hazard, where discrimination is partly a learned local pattern rather than a universal physical signature, is an open empirical question.

\subsection{Temporal stability and data drift}

OpenStreetMap, Overture, and Google Places are continuously updated, incompletely mapped, and subject to commercial biases. Whether a model trained on 2024 POI data performs equivalently on 2027 data is untested. Required follow-ons: temporal retrospective scoring the same grid across multiple POI snapshots; geographic-coverage audit against a published POI completeness metric; incremental retraining policies that respond to drift. \textit{For the broader class:} drift sensitivity is class-wide. Any informal-hazard methodology built on continuously-updated public POI databases inherits these temporal artefacts.

\subsection{Partnerships with community-mapping initiatives}
\label{sec:roadmap-community}
Several urban-deprivation mapping initiatives in LMICs are now generating fine-grained, community-validated layers (road-access deprivation, accumulated waste-piles, water and emergency-service access, and similar variables) at neighbourhood scale. These layers were not used as features in this paper, but they cover exactly the kind of structural and infrastructural context in which informal recycling clusters tend to operate. Co-developing feature banks with such initiatives, and where possible co-collecting labels through their existing community-mapping networks, would address the two most binding constraints of the methodology in one move: the POI-coverage gap in deprived urban areas, and the absence of independent positive-class inventories outside Pure Earth's TSIP. 

\subsection{Generalisation beyond ULAB}
\label{sec:roadmap-beyond}

The four hardness characteristics in Section~\ref{sec:hard} define a class beyond ULAB: informal e-waste burning, household-scale ASGM, brick kilns, illegal dumping, and small tanneries share sub-resolution operational scale, residential embeddedness, no distinctive physical signature, and absence from formal registries. The methodology is in principle applicable to each, with appropriate POI lexica (e.g. ``sluice'', ``amalgam'' for ASGM; ``tannery'', ``chrome'' for hide processing), local-language transliterations, and analogous socioeconomic priors. ULAB is a useful first test case because TSIP gives it the largest labelled-positive set; whether the same methodological challenges recur in the other hazards is the open empirical question raised in Section~\ref{sec:roadmap}.

\section{Conclusion}

Environmental satellite remote sensing has solved much of the easier half of the spatial-epidemiology localisation problem: hazards with direct, spatially-resolvable physical signatures larger than the sensor footprint can now be located, classified, and monitored from space at near-global scale. The harder half (informal, residentially-embedded, sub-resolution operations) remains structurally invisible to those methods, and includes some of the largest unaddressed sources of environmental-health burden in LMICs. This paper has proposed contextual geospatial features as a methodological alternative for this hidden half, and stress-tested the proposal on informal ULAB recycling, a high-stakes instance with sufficient field-verified labels to make quantitative evaluation tractable.

The empirical findings are mixed and informative. The methodology finds important statistical signals across multiple validation tiers; training-set rank discrimination, deployment-region calibration tightening, area-based regional enrichment around documented informal sites, out-of-region TSIP transfer (highly significant in non-NCR India), and a fully out-of-source validation against 131 regulatory-confirmed formal recyclers; but training-set AUC alone does not statistically distinguish the engineered features from a simple socio-demographic baseline, and so they could be easily missed by researchers who don't unpack the problem further. These observations carry directly to the broader class of informal-hazard detection problems.

The larger contribution of this paper is the structured articulation of what remains unsolved. Section~\ref{sec:roadmap} sets out the open problems that would help convert a methods-diagnosis paper like this one into a methods-solution paper, that can be used by practitioners. These directions are not best pursued in isolation, they depend on partnerships with field organisations that can enable wider ground truth, with community-mapping initiatives, and with the broader geospatial machine-learning community that has tools (multilingual embeddings, foundation models, spatial cross-validation frameworks) that we have not exploited here. The environmental-health and geospatial machine-learning communities are invited to treat informal-hazard detection from public contextual features as a shared challenge problem: its combination of extreme label scarcity, geographic heterogeneity, and very low field-detection rates makes it a genuine test bed for methods claiming to operate under real-world LMIC constraints. 

The methodology proposed here is also a starting point rather than an endpoint. Each new informal environmental hazard type will require its own POI lexicon, its own consultation with field-investigation experts, and its own validation tier built against whatever labelled inventory exists. The benefit of the approach is that the bones generalise (channel structure, evaluation strategy, the deployment-vs-AUC framing, etc.) even when the feature design has to be redone. The hidden half of the spatial-epidemiology localisation problem remains hidden. Reliably finding it, and deploying predictions ethically, is an unsolved challenge worth working on. 

\section*{Acknowledgment}

The lead author thanks David R. Boyd for his guidance and presence throughout this work, and for situating the project in the broader environmental-health and human-rights perspective that underwrites its public-health framing. She also thanks Pure Earth for the Toxic Sites Identification Programme (TSIP) database, which is the source of the field-verified ULAB labels used throughout this paper. Finally, she thanks Michael Brauer (IHME), Bruce Lanphear (Simon Fraser University), and Lee Crawfurd (CGD) for agreeing to meet, for allowing her to present early versions of this work, and for the feedback that informed the paper's eventual emphasis.

{\footnotesize
\bibliographystyle{IEEEtran}
\bibliography{refs}
}

\appendix
\renewcommand{\thesection}{\Alph{section}}
\renewcommand{\thesubsection}{\Alph{section}.\arabic{subsection}}

\section{Cross-Validation, Permutation, and Transfer Details}
\label{sec:appx-cv}

Per-fold ROC-AUC and F1 for the Final model (n = 164, 5-fold stratified, random\_state = 42) are listed in Table~\ref{tabIII}: per-fold AUC 0.687, 0.700, 0.878, 0.830, 0.759; mean $\pm$ std AUC 0.771 $\pm$ 0.083; per-fold F1 at $\tau = 0.5$ are 0.250, 0.500, 0.667, 0.667, 0.545; mean $\pm$ std 0.526 $\pm$ 0.171 (the F1 instability across folds is itself indicative of the small-sample regime, see Section~\ref{sec:results-cv}). Pooled 5-fold CV AUC point estimates under each configuration: Open-data baseline 0.718; Full feature bank 0.749; Final model 0.765; 2-feature socio-demographic 0.759. These point estimates lie within fold-to-fold variability (Table~\ref{tabII}) and are not statistically distinguishable on the training-set 5-fold CV AUC. Country-stratified label permutation (300 shuffles, labels permuted within country, which preserves the within-country class structure): null mean AUC 0.541 (SD 0.070); observed AUC 0.765; empirical $p < 0.001$. The observed AUC sits well outside the country-preserved null distribution, so the training-set signal is not an artefact of country composition. Cross-country LOCO transfer (Section~\ref{sec:results-transfer}): BGD$\to$IND AUC = 0.672 (n$_{\text{test}}$ = 86, 17 pos); IND$\to$BGD AUC = 0.609 (n$_{\text{test}}$ = 78, 33 pos). Both directions sit modestly above chance, well below the training-set 5-fold CV AUC of 0.765 on the full training set, consistent with within-country structure that does not transfer cleanly across the BGD/IND boundary.

\begin{table}[!t]
\caption{TABLE VII. PER-FOLD ROC-AUC AND F1 (Final model, n = 164)}
\label{tabIII}
\centering
\setlength{\tabcolsep}{3pt}
\footnotesize
\begin{tabular}{@{}lrrrrr@{}}
\hline
\textbf{Fold} & \textbf{1} & \textbf{2} & \textbf{3} & \textbf{4} & \textbf{5} \\
\hline
AUC & 0.687 & 0.700 & 0.878 & 0.830 & 0.759 \\
F1 @ 0.5 & 0.250 & 0.500 & 0.667 & 0.667 & 0.545 \\
n test & 33 & 33 & 33 & 33 & 32 \\
\hline
\end{tabular}
\end{table}

\section{Coverage-Robustness and Moran's I Details}
\label{sec:appx-coverage}

Coverage-tertile-stratified AUC across the labelled training pool (Places-per-capita tertiles): low coverage AUC $\approx$ 0.68, middle $\approx$ 0.58, high $\approx$ 0.80. NCR fraction-above-threshold by Google Places coverage bin (under the Final model): 0 places nearby (3,192 cells, 85 \% of NCR) flag at 12.7 \%; 1--5 places (1 cell) 0.0 \%; 6--15 places (0 cells in NCR scan) n/a; 16--20 places (553 cells, 15 \%) flag at 58.0 \%. Inverse-coverage-weighted (weights $\propto$ 1/(places + 1)) NCR fraction p $\geq$ 0.5: 13.0 \% vs 19.4 \% unweighted. Toxics-Link neighbourhood test under inverse-coverage weights: the TSIP-neighbourhood mean score sits well above the inverse-coverage-weighted random-NCR null (Mann--Whitney $p < 0.0001$ vs $p = 0.0002$ under uniform weights, reported as exploratory; the per-cell statistics share spatial autocorrelation, see Table~\ref{tabMoran}). The test strengthens under inverse-coverage weighting because down-weighting high-coverage controls makes the random-control distribution lower-scoring while leaving the TSIP-neighbourhood means roughly unchanged. Moran's I on the Final-model NCR scoring with k = 8 nearest-neighbour weights is reported in Table~\ref{tabMoran}: the score field is meaningfully more spatially structured than random noise (Moran's I $\approx$ 0) but less clumped than population or raw industrial counts.

\begin{table*}[!t]
\caption{TABLE VIII. MORAN'S I ON 3,746-CELL NCR GRID (FINAL MODEL), k = 8 NEAREST-NEIGHBOUR SPATIAL WEIGHTS}
\label{tabMoran}
\centering
\setlength{\tabcolsep}{4pt}
\footnotesize
\begin{tabular}{p{6cm}rrr}
\hline
\textbf{Variable} & \textbf{Moran's I} & \textbf{Null mean} & \textbf{p-value} \\
\hline
Final model (predicted scores) & +0.292 & $-$0.001 & < 0.001 \\
Open-data baseline (predicted scores) & +0.266 & $-$0.001 & < 0.001 \\
pop\_density (baseline reference) & +0.794 & $-$0.000 & < 0.001 \\
industrial\_count\_1km (baseline reference) & +0.823 & $-$0.001 & < 0.001 \\
\hline
\end{tabular}
\end{table*}

\section{Feature-Extraction Pipeline}
\label{sec:appx-extraction}
The feature vector for each candidate location is assembled from five public
data sources sampled at fixed metric buffers; lat/lon offsets are adjusted
for the local cosine so a ``500\,m radius'' is 500\,m everywhere we deploy.
Socio-demographic priors carry the broadest spatial signal, open POI sources
add infrastructure and industrial context at neighbourhood scale, and the
multi-channel Google Places features encode the case-specific ULAB ecology.

\textit{Socio-demographic.} The population feature is the WorldPop 2020
raster sampled at the candidate point at its native $\sim$100\,m resolution.
The wealth feature is the Meta Relative Wealth Index at the nearest
country-level grid point, with each RWI value covering a $\sim$2.4\,km cell.
Country selection follows the candidate's reported country. Both surfaces
are themselves machine-learning products, a point we return to throughout
Section~\ref{sec:results} when interpreting attribution.

\textit{Open POI (OSM and Overture).} From OpenStreetMap we keep road and
waterway proximity, the metric distance from the candidate to the nearest
road or watercourse within 1\,km. From Overture Maps we extract an
industrial-POI density panel: POIs within 500\,m and 1\,km whose category or
name matches a fixed keyword union covering battery, recycling, scrap,
metal-work, foundry, auto-repair, motorcycle, and similar industrial
categories. The 500\,m count is broken down into five sub-categories
(battery, metal, auto, scrap, electrical) using a first-match rule, and a
count of distinct non-zero sub-categories serves as a category-diversity
feature. Two binary flags derived from Overture's primary tag mark whether
the candidate sits on an industrially-tagged record and whether that tag is
one of the higher-risk ULAB-adjacent categories.

\textit{Google Places multi-channel.} This is the engineered backbone. For
each candidate we query the Google Places Nearby Search API at a 500\,m
radius and cache the response on disk so the same point is never queried
twice; the pipeline is therefore reproducible at a fixed cost. Each cached
POI carries multiple text channels (display name, primary category, full
type list, formatted address), and we match each channel against a
five-tier keyword lexicon of English, Hindi, and (where available) Bengali
ULAB-adjacent terms, ordered roughly by directness: battery, smelting and
recycling, metal-work, auto-repair, electrical. A short anti-keyword list
(laptop, phone, restaurant, school, and similar) suppresses common
confusables in display names. The per-channel keyword counts are aggregated
into a composite ULAB score that weights each tier by directness, with
battery-direct hits dominating, electrical-tier hits entering only weakly,
and anti-keyword hits subtracting. Structural place-density features (total
cached places, address-token hits for ``industrial area / estate / zone /
park'', and primary-type densities for auto-parts stores, car-repair shops,
tire shops, and manufacturers) capture the surrounding commercial ecology
even when no keyword tier fires. Two binary indicators flag whether the
candidate has at least one direct ULAB-keyword match and whether its
primary types include the strongest industrial categories.

The complete feature schema, including column names, exact tier weights,
and the keyword lexicon, is in the public code repository released with the
paper \cite{ref23}.

\section{Neighbourhood-Validation Details}
\label{sec:appx-nbhd}

\begin{figure}[!t]
\centering
\includegraphics[width=\columnwidth,keepaspectratio]{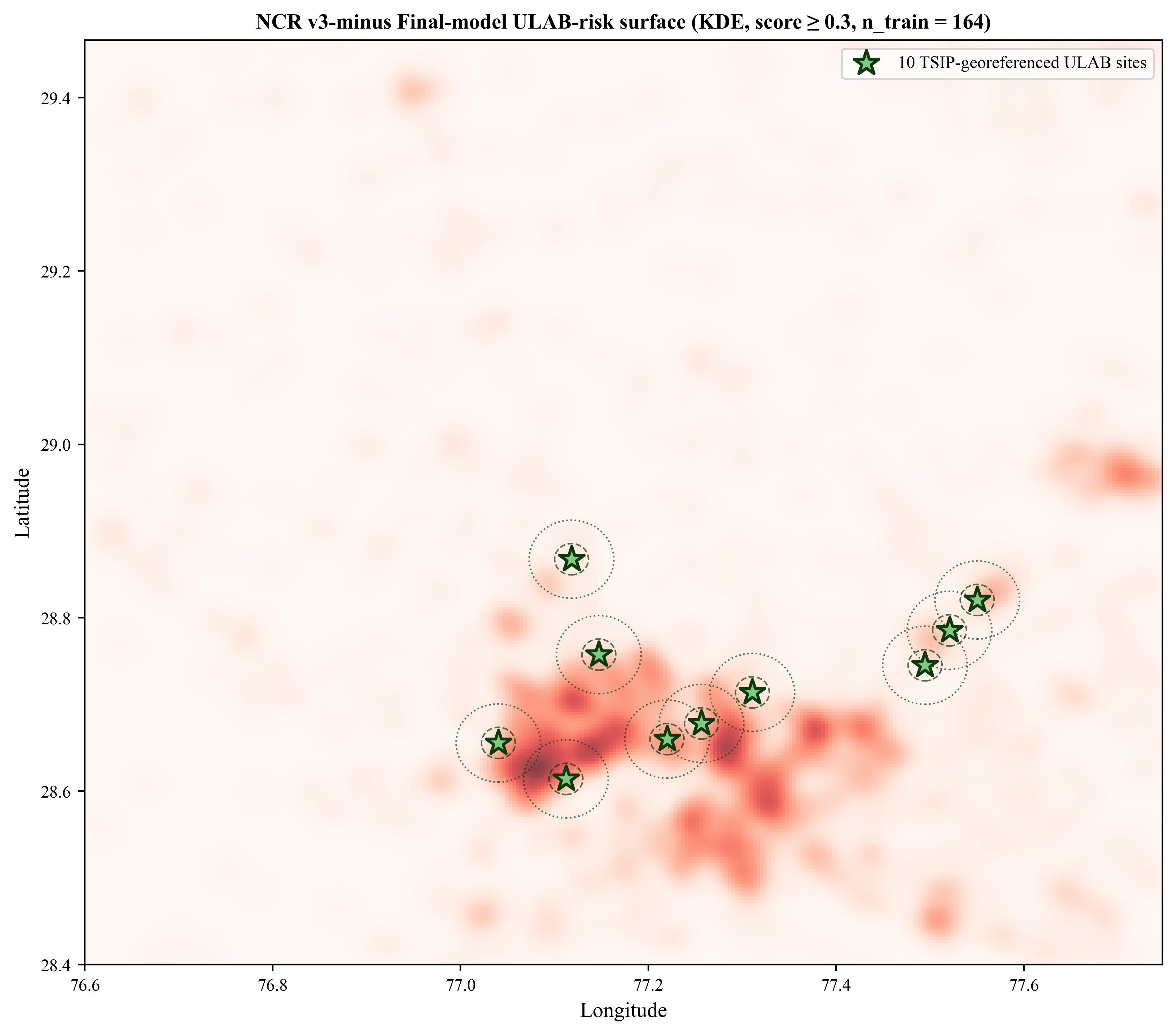}
\caption{NCR Final-model ULAB-risk surface. Score-weighted 2-D kernel density estimate (Gaussian kernel, bandwidth $\approx$ 3\,km) of grid cells with Final-model score $\geq 0.3$, normalised so the maximum density is 1. Hot-spots concentrate in central Delhi, Ghaziabad, the Noida--Greater Noida corridor, and along the Muzaffarnagar industrial belt to the north. Green stars mark the 10 TSIP-georeferenced ULAB sites used for external validation, with 2\,km and 5\,km buffer rings.}
\label{fig5}
\end{figure}

\begin{figure}[!t]
\centering
\includegraphics[width=\columnwidth,keepaspectratio]{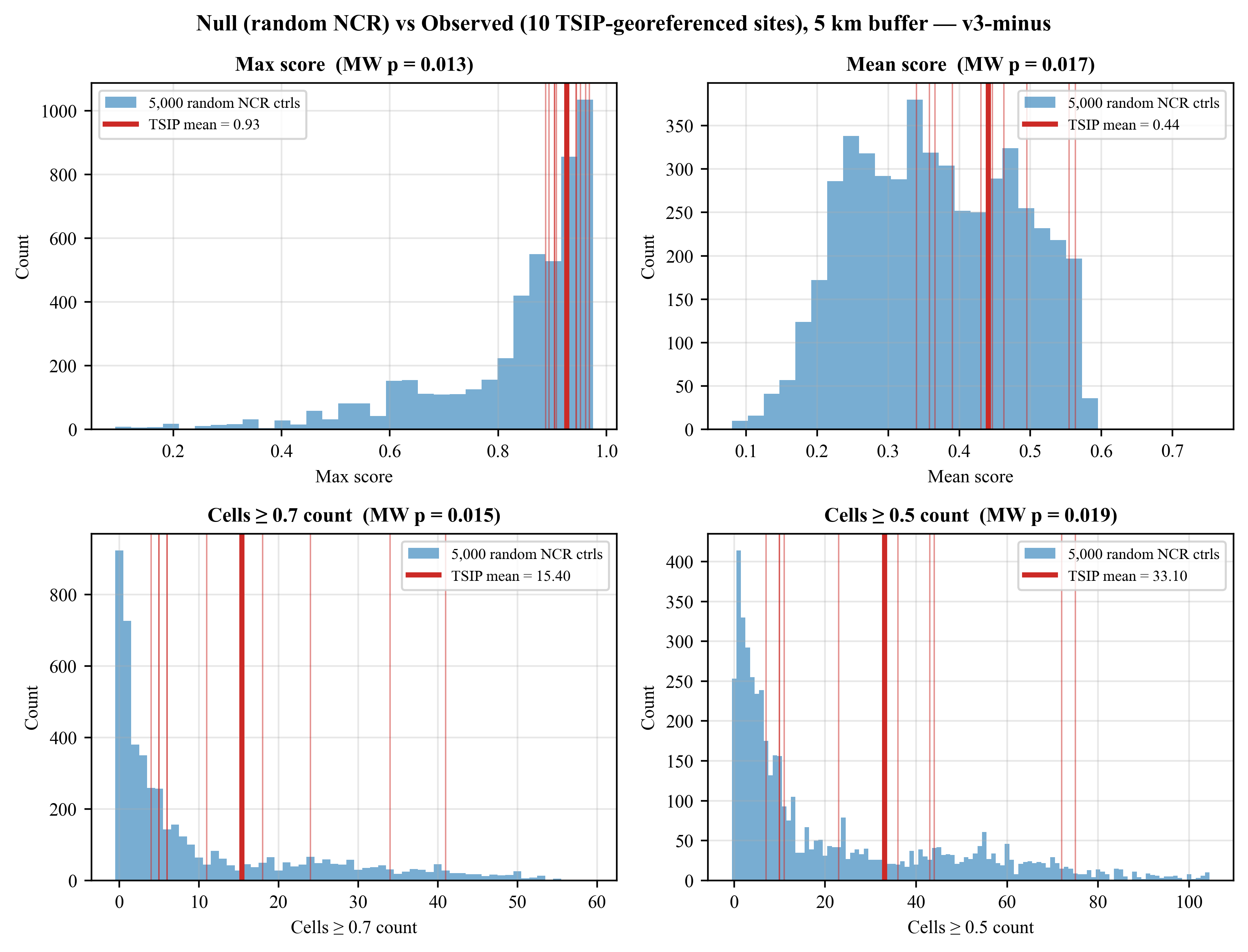}
\caption{Null (histogram, random NCR neighbourhoods) vs.\ observed (red vertical lines, 10 TSIP-georeferenced ULAB neighbourhoods) distributions of four neighbourhood statistics at 5\,km buffer radius. The bold red line is the TSIP-site mean. Reported as exploratory; see caveat in Appendix~\ref{sec:appx-nbhd}.}
\label{fig6}
\end{figure}

This appendix collects the per-buffer numerical breakdown and supporting figures for the area-based NCR neighbourhood test (Section~\ref{sec:results-nbhd}). Table~\ref{tabNbhd} reports max-score and high-confidence-count statistics at three buffer radii against a 10{,}000-sample random-NCR null; Fig.~\ref{fig5} shows the NCR Final-model risk surface as a score-weighted kernel density estimate with the 10 TSIP buffer rings overlaid; Fig.~\ref{fig6} shows the null vs observed distributions at the 5\,km buffer for four neighbourhood statistics. We report these as supplementary evidence supporting the regional-but-not-site-level claim made in Section~\ref{sec:results-nbhd}. The $n = 10$ TSIP-NCR comparison is severely underpowered and the per-cell statistics share strong spatial autocorrelation (Table~\ref{tabMoran}); the reported $p$-values should be read as exploratory ranking indicators, not as confirmatory tests.

\begin{table}[!t]
\caption{TABLE IX. NEIGHBOURHOOD VALIDATION, FINAL MODEL POST-HARD-NEGATIVE (10 TSIP-NCR sites vs 10{,}000 random-NCR neighbourhoods). Reported as exploratory; see caveat above and Section~\ref{sec:results-nbhd}.}
\label{tabNbhd}
\centering
\setlength{\tabcolsep}{3pt}
\footnotesize
\begin{tabular}{@{}llrrr@{}}
\hline
\textbf{Buffer} & \textbf{Metric} & \textbf{TL mean} & \textbf{Null mean} & \textbf{p} \\
\hline
1 km & max score & 0.517 & 0.529 & 0.557 \\
1 km & count $\geq$ 0.7 & 0.50 & 0.68 & 0.670 \\
2 km & max score & 0.777 & 0.662 & 0.068 \\
2 km & count $\geq$ 0.7 & 1.40 & 2.18 & 0.753 \\
5 km & max score & 0.927 & 0.826 & \textbf{0.004} \\
5 km & count $\geq$ 0.7 & 15.4 & 10.3 & 0.126 \\
\hline
\end{tabular}
\end{table}

\section{Cross-Population Summary of Feature Importance and Ablation}
\label{sec:appx-summary}

\begin{figure*}[!t]
\centering
\includegraphics[width=\textwidth,keepaspectratio]{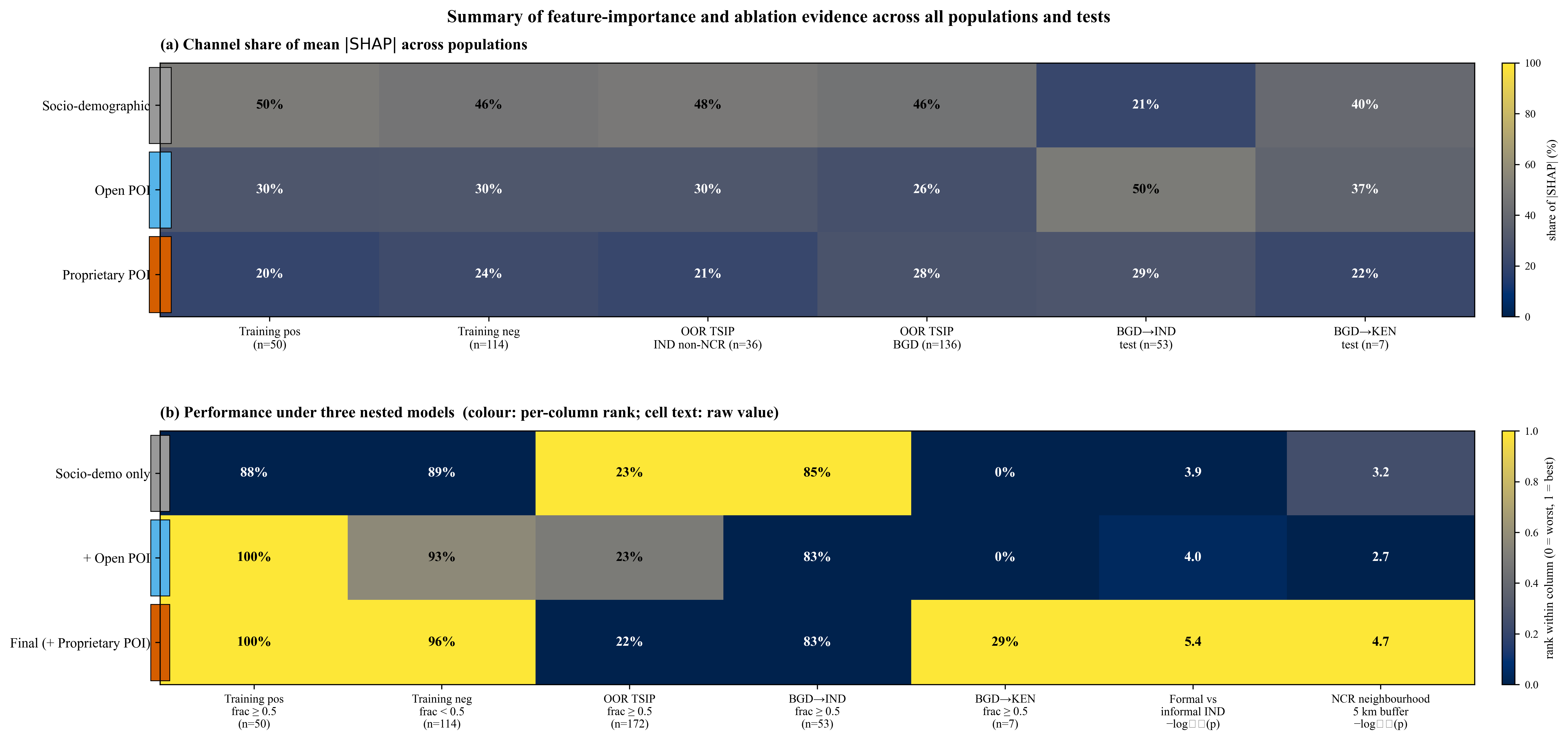}
\caption{Summary of feature-importance and ablation evidence across all populations and tests. \textbf{(a)} Channel share of mean $|\mathrm{SHAP}|$ across six populations (training positives, training negatives, OOR TSIP IND non-NCR, OOR TSIP BGD, BGD-trained at IND test, BGD-trained at KEN test); cell values sum to 100\,\% per column. Shares are computed over the full 20-feature panel (cf.\ Figs.~\ref{fig9}(a), \ref{fig11}(a) which restrict to the top-10 features, hence small numerical differences). Coloured row badges indicate the channel: grey = socio-demographic, sky blue = Open POI, vermillion = Proprietary POI. The Proprietary-POI share is higher at every out-of-training population than at the training set itself, with the largest shares at the Bangladesh OOR TSIP cohort and the BGD-trained model's IND test. \textbf{(b)} Performance heatmap under three nested models (socio-demographic only, $+$ Open POI, Final $+$ Proprietary POI) across seven tests; cell colour is per-column rank (cividis: dark = worst, light = best within column), cell text is the raw metric. Tests, left to right: frac $\geq 0.5$ at training positives, frac $< 0.5$ at training negatives, frac $\geq 0.5$ at the 172 OOR TSIP positives, frac $\geq 0.5$ at the 53 BGD$\to$IND test positives, frac $\geq 0.5$ at the 7 BGD$\to$KEN test positives, and $-\log_{10}(p)$ for the two Mann--Whitney validation hypothesis tests of Section~\ref{sec:results-nbhd}.}
\label{fig15}
\end{figure*}

Fig.~\ref{fig15} collects the cross-population SHAP-attribution and per-test performance evidence reported across Sections~\ref{sec:results-fi}, \ref{sec:results-transfer}, and \ref{sec:results-nbhd} into a single two-panel heatmap. Panel (a) shows the channel share of mean $|\mathrm{SHAP}|$ across six populations (training positives, training negatives, OOR TSIP India non-NCR, OOR TSIP Bangladesh, BGD-trained model at IND test, BGD-trained model at KEN test); shares are computed over the full 20-feature panel (rather than restricted to the top-10 features as in Figs.~\ref{fig9}(a) and \ref{fig11}(a)), so the per-population numbers here are quantitatively distinct from the top-10 channel shares quoted in Section~\ref{sec:results-fi} prose. The substantive pattern is the same in both views: the Proprietary-POI share is higher at every out-of-training population than at the training set itself. Panel (b) shows performance under three nested models across seven tests; cell colour is per-column rank (cividis: dark = worst, light = best within column), cell text is the raw metric (frac $\geq \tau$ or $-\log_{10}(p)$ depending on the test).

Table~\ref{tabNested} reports the full panel of one-sided Mann--Whitney $p$-values that underlies the heatmap in Fig.~\ref{fig15}(b). As noted in Section~\ref{sec:results-nbhd}, these values should be read as exploratory ranking indicators of channel contribution: nearby urban cells share POI context and the underlying score field is spatially autocorrelated (Table~\ref{tabMoran}), so Mann--Whitney's independence assumption is violated. The substantive content is the direction and magnitude of the change as channels are added, not the absolute values.

\begin{table*}[!t]
\caption{TABLE X. NESTED-MODEL BREAKDOWN OF VALIDATION MANN--WHITNEY $p$-VALUES (three nested models on the same n = 164 training set; $p$-values one-sided in the direction stated). The third column is the Final model, which adds the Proprietary-POI channel on top of the $+$ Open POI panel. Reported as exploratory indicators only; see Section~\ref{sec:results-nbhd} and the caveat above for the independence-assumption violation.}
\label{tabNested}
\centering
\setlength{\tabcolsep}{5pt}
\footnotesize
\begin{tabular}{p{8.0cm}ccc}
\hline
\textbf{Hypothesis} & \textbf{Socio-demo only} & \textbf{+ Open POI} & \textbf{Final (+ Proprietary POI)} \\
\hline
H2: Informal IND non-NCR $>$ formal IND non-NCR (36 vs 100) & $1.2{\times}10^{-4}$ & $1.0{\times}10^{-4}$ & $3.7{\times}10^{-6}$ \\
H3: All TSIP-informal $>$ all formal (172 vs 131) & $0.031$ & $0.217$ & $0.059$ \\
H4: TSIP-informal IND non-NCR $>$ random urban IND (36 vs 557) & $1.6{\times}10^{-4}$ & $5.1{\times}10^{-6}$ & $3.0{\times}10^{-8}$ \\
H5: TSIP-informal BGD $>$ random urban BGD (136 vs 600) & $0.982$ & $0.077$ & $0.028$ \\
H6: NCR 5 km max-score $>$ random null (10 sites vs 10{,}000) & $6.5{\times}10^{-4}$ & $2.1{\times}10^{-3}$ & $1.9{\times}10^{-5}$ \\
\hline
\end{tabular}
\end{table*}

\section{Data Availability}
\label{sec:appx-data}

Upon publication, the public GitHub repository for this work \cite{ref23} will release the code and all publicly-available data sources used in the analysis (subject to upstream licences). Pure Earth's TSIP database is the source of training positives and out-of-region validation positives, accessed under their research-use terms. Indian regulatory registries used for Section~\ref{sec:results-nbhd} formal-vs-informal validation are public PDFs at cpcb.nic.in, mpcb.gov.in, dpcc.delhigovt.nic.in, and tnpcb.gov.in (last accessed April 2026). Google Places features were extracted under the Google Maps Platform terms of service and the cached responses are not redistributed; reproducing the Google-Places-dependent configurations therefore requires a Google Maps Platform API key. WorldPop population grids and Meta Relative Wealth Index data are CC-BY-licensed and publicly downloadable. OpenStreetMap and Overture Maps data are ODbL-licensed and publicly accessible.

\end{document}